\documentclass[11pt,a4paper]{article}
\pdfoutput=1
\usepackage{jheppub}

\usepackage{bbm}
\usepackage{slashed}
\usepackage{mathtools}
\usepackage[matrix,arrow]{xy}
\usepackage{float}
\usepackage{upgreek}

\newcommand{\qed}{\nobreak \ifvmode \relax \else
      \ifdim\lastskip<1.5em \hskip-\lastskip
      \hskip1.5em plus0em minus0.5em \fi \nobreak
      \vrule height0.75em width0.5em depth0.25em\fi}

\DeclareMathAlphabet{\mathpzc}{OT1}{pzc}{m}{it}
\def\tHooft{\mbox{'t Hooft }}

\def\NN{\mathcal{N}}

\def\AA{\mathcal{A}}
\def\CC{\mathcal{C}}
\def\EE{\mathcal{E}}

\def\LL{\mathcal{L}}
\def\AA{\mathcal{A}}

\def\MM{\mathcal{M}}
\def\GG{\mathcal{G}}

\def\UU{\mathcal{U}}

\def\Wbar{\overline{W}}

\def\fMM{\overline{\underline{\MM}}}
\def\gm{\gamma_{\rm m}}

\def\pd{\partial}
\def\ed{  \textrm{d}}
\def\eD{ \textrm{D}}
\def\rnk{ \, \textrm{rnk} \, }

\def\ad{ \, \textrm{ad} }

\def\tr{ \, \textrm{tr} }
\def\ind{ \, \textrm{ind}  }

\def\diag{ \, \textrm{diag}}
\def\sgn{ \, \textrm{sgn} }
\def\vol{ \, \textrm{vol} }
\def\half{\frac{1}{2}}

\def\ie{{\it i.e.}}
\def\eg{{\it e.g.}}

\def\etc{{\it etc}}

\def\be{\begin{equation}}
\def\ee{\end{equation}}
\def\bea{\begin{eqnarray}}
\def\eea{\end{eqnarray}}

\title{Brane bending and monopole moduli}
\author[a]{Gregory W.~Moore,}
\author[b]{Andrew B.~Royston,}
\author[c]{Dieter Van den Bleeken}

\affiliation[a]{NHETC and
Department of Physics and Astronomy, Rutgers University \\
126 Frelinghuysen Rd., Piscataway NJ 08855, USA}
\affiliation[b]{George P.\ \& Cynthia Woods Mitchell Institute for Fundamental Physics and Astronomy, \\
Texas A\&M University, College Station, TX 77843, USA}
\affiliation[c]{Physics Department, Bo\u{g}azi\c{c}i University\\
 34342 Bebek / Istanbul, TURKEY}

\emailAdd{gmoore@physics.rutgers.edu}
\emailAdd{aroyston@physics.tamu.edu}
\emailAdd{dieter.van@boun.edu.tr}

\abstract{We study intersecting brane systems that realize a class of singular monopole configurations in four-dimensional Yang--Mills--Higgs theory.  Singular monopoles are solutions to the Bogomolny equation on $\mathbb{R}^3$ with a prescribed number of singularities corresponding to the insertion of 't Hooft defects.   We use the brane construction to motivate a recent conjecture on the conditions for which the moduli space of solutions is non-empty.  We also show how branes provide physical intuition for various aspects of the dimension formula derived in \cite{MRVdimP1}, including the contribution to the dimension from the defects and its invariance under Weyl reflections of the 't Hooft charges.  Along the way we uncover and illustrate new dynamical phenomena for the brane systems, including a description of smooth monopole extraction and bubbling from 't Hooft defects.}


\begin{document}
\begin{flushright} MIFPA-14-14 \end{flushright}
\maketitle

\section{Introduction and summary}

Intersecting brane systems in string theory provide one with a great deal of intuition about field theory phenomena.  The basic idea is to realize the field theory of interest as the low energy description of the degrees of freedom on a brane worldvolume.  Then many field theory features such as moduli spaces of vacua, internal symmetries, and solitonic particles can be understood in terms of geometric properties of the brane system in the remaining spatial directions orthogonal to the worldvolume.  Oftentimes a given field theory can be engineered in more than one way, leading to additional insights.  See \eg\ \cite{Giveon:1998sr} for an extensive review on the engineering and analysis of gauge theories via branes.

In this paper we consider the simplest possible intersecting brane system that realizes singular monopole configurations in $1+3$-dimensional gauge theory.  The gauge theory is obtained as the low energy description of a system of $N$ parallel D3-branes.  Each smooth fundamental monopole is represented by a finite length D1-string stretched between consecutive D3-branes, each of which can move in three directions and has a fourth modulus dual to turning on electric charge.  Studying this system from the D1-string point of view leads to  the ADHM--N description of monopoles through Nahm's equation \cite{Diaconescu:1996rk}, but we will not make use of this result here.

Taking the brane picture as motivation, we obtain a prescription for constructing singular monopole field configurations as a certain limit of smooth monopole configurations in a higher rank gauge group.  In the context of branes this limit corresponds to making some of the D1-strings semi-infinite, an idea that was first pointed out in a T-dual context in \cite{Hanany:1996ie}.\footnote{This in turn motivates the study of singular monopoles and their moduli spaces via the Nahm equation on a semi-infinite interval.  Explicit results for singular $\mathfrak{su}(2)$ monopole moduli spaces have been obtained in this way in \cite{Cherkis:1997aa,Cherkis:1998xca,Cherkis:1998hi,Cherkis:2007qa}.}  Our prescription involves a projection of the Lie algebra-valued fields to a sub-algebra, followed by a limit on the Higgs vev.

Our goal is to understand and interpret the dimension formula derived in \cite{MRVdimP1} in the context of these brane systems.  This formula gives the dimension of the moduli space of solutions to the Bogomolny equation, $F = \star \eD \Phi$, on oriented Euclidian three-space with some number $N_t$ of points removed, $\mathbb{R}^3 \setminus \{ \vec{x}_n \}_{n=1}^{N_t}$, subject to certain boundary conditions at the $\vec{x}_n$ and at infinity.  The pertinent details of this result will be reviewed below in section \ref{sec:review}.  Here we simply note that the dimension of this space depends on three types of data, all of which are simultaneously valued in a Cartan subalgebra $\mathfrak{t} \subset \mathfrak{g}$.  These are the magnetic charge and asymptotic Higgs vev, $(\gm ; \Phi_\infty)$, and the \tHooft charges $P_n$.  We henceforth refer to $\mathfrak{t}$ as \emph{the} Cartan subalgebra.  We denote the moduli space as $\fMM = \fMM \left((\vec{x}_n, P_n)_{n=1}^{N_t} ; \gm; \Phi_\infty \right)$.\footnote{The moduli space itself, as a Riemannian manifold, will depend on both the locations and charges of \tHooft defects.  The dimension, however, only depends on the charges.}  We restrict to the case of regular $\Phi_\infty$, corresponding to maximal symmetry breaking where the gauge group $G$ is broken to the Cartan torus $T$.  More precisely, the dimension formula only depends on the Weyl orbit, $[P_n]$ of $P_n$; it gives the same value for any pair $P_n, P_n'$ related by a Weyl transformation.  This is expected, since Weyl transformations of an \tHooft charge can be implemented by local gauge transformations and hence it is only the Weyl orbit that is physical.  

We want to interpret the dimension of the moduli space as four times the number of mobile, finite-length D1-strings, just as one can do in the case of smooth monopoles.  We find that the simplistic picture of semi-infinite D1-strings ending on D3-branes is sufficient for this purpose in some simple examples, but for \tHooft defects with generic charges this naive counting gives the wrong answer.  We show how the apparent discrepancy is resolved when the more precise picture of brane bending is taken into account, where semi-infinite D1-strings are replaced with a ``BIon'' spike \cite{Gibbons:1997xz,Callan:1997kz}.

The bent brane representation of \tHooft defects provides insight into why the naive counting of D1-string segments fails.  One finds that there is an important distinction between 't Hooft defects for which the BIon spikes of different D3-branes intersect each other and defects for which they do not.  In the former case, the intersection of the spikes can be interpreted as the presence of some number of smooth monopoles centered on the defect.  Allowing all such monopoles to move off the defect results in a new configuration of the latter type, where the BIon spikes do not intersect.  This configuration represents an \tHooft defect whose charge has changed by a Weyl transformation.  Since the new \tHooft charge is physically equivalent to the original one, and since the asymptotic data does not change, this process, which we refer to as monopole extraction, can be viewed as motion on a fixed moduli space.

\tHooft charges for which the corresponding BIon spike has no brane intersections have the property of being in the closure of the anti-fundamental Weyl chamber.  Here the notion of positive roots, $\alpha \in \Delta^+$, used to define this chamber, is determined from the asymptotic Higgs field: $\alpha \in \Delta^+ \iff \langle \alpha ,\Phi_\infty \rangle > 0$.  We denote the representative of $[P_n]$ in the closure of the anti-fundamental chamber by $P_{n}^-$.  It has the property $\langle \alpha, P_{n}^- \rangle \leq 0$, for all $\alpha \in \Delta^+$, and it plays a special role.\footnote{Similar observations were made in the study of supersymmetric boundary conditions for $\NN = 4$ SYM on the half-space \cite{Gaiotto:2008sa}, where a closely related brane set-up was used, and the relevance of the BIon spike picture was also noted.}  Using it, we define the \emph{relative magnetic charge} $\tilde{\gamma}_{\rm m} := \gm - \sum_n P_{n}^-$.  One can argue \cite{MRVdimP1} that this quantity sits in the co-root lattice $\Lambda_{\rm cr} \subset \mathfrak{t}$, and our conjecture for when the moduli space $\fMM$ is non-empty is that $\tilde{\gamma}_{\rm m}$ is a non-negative linear combination of simple co-roots: $\tilde{\gamma}_{\rm m} = \sum_{I=1}^{\rnk{\mathfrak{g}}} \tilde{m}^I H_I$, with $\tilde{m}^I \geq 0$, $\forall I$.  Thus, in analogy with \cite{Weinberg:1979zt}, we interpret the $\tilde{m}^I$ as the number of smooth fundamental monopoles of type $I$, in the presence of the defects $P_n$.  Indeed, our dimension formula takes the simple form $\sum_I 4 \tilde{m}^I$ in terms of these coefficients.  These numbers are also the numbers of D1-string segments in the brane setup, once the process of monopole extraction has been carried out.\footnote{One may wonder why we do not simply restrict consideration to singular monopole configurations where all \tHooft charges satisfy $P_n = P_{n}^-$, since these are physically equivalent charges.  While the charges are equivalent, two field configurations in which we simply exchange $P_n$ for $P_{n}^-$ are \emph{not} physically equivalent.  Rather one must make a local gauge transformation that acts as a Weyl transformation on $P_n$ at $\vec{x} = \vec{x}_n$, while going to the identity at infinity.  Such a gauge transformation might produce a field configuration that is less convenient to work with.  The study of the Cartan-valued solutions in \cite{MRVdimP1} provides an excellent example of this.}

We distinguish the process of monopole extraction, where the Weyl orbit $[P_n]$ does not change, from that of monopole bubbling \cite{Kapustin:2006pk}, where the \tHooft defect emits or absorbs a smooth monopole, such that the Weyl orbit does change.  We show how monopole bubbling can also be understood in terms of brane motion, in some simple examples.  This leads to some discussion about the structure of possible singular loci of the moduli space $\fMM$.

Finally, the dimension formula exhibits interesting jumping behavior as the Higgs vev $\Phi_{\infty}$ is varied.  In fact, in \cite{MRVdimP1} we derived a more general formula for the dimension of a certain vector bundle over $\fMM$.  This is the index bundle of the Dirac operator constructed from the singular monopole configuration \cite{Taubes:1984je,Manton:1993aa}, acting on matter fermions in a representation $\rho$ of $G$.  The fiber over a point on moduli space representing a singular monopole configuration is identified with the kernel of the Dirac operator; (there is a vanishing theorem for the kernel of the adjoint operator).  When $\rho$ is the adjoint representation, the vector bundle is related to the tangent bundle, and twice its dimension gives the dimension of $\fMM$.  We give a brane interpretation of the jumping phenomena of the index in the case of the adjoint representation and the fundamental representation.  For the fundamental representation this involves the addition of a ``flavor'' D7-brane.  The matter fermions are realized as low energy degrees of freedom of strings stretched between the D7-brane and the D3-branes.

A brief outline of the paper is as follows.  In section \ref{sec:review} we define the singular monopole moduli space $\fMM$ and recall necessary results from \cite{MRVdimP1}.  We refer the reader to that paper for derivations, as well as a summary of previous results in the literature on singular monopoles and motivation for their study.  In section \ref{sec:branereview} we briefly recall how smooth monopoles are represented as configurations of D3-branes and D1-strings.  In section \ref{sec:sm2si} we describe the projection and limiting procedure for producing singular monopole configurations from smooth ones, and interpret it in terms of brane motion.  In sections \ref{sec:su3} and \ref{sec:suN} we apply the procedure to some simple examples, first for $\mathfrak{g} = \mathfrak{su}(3)$ and then $\mathfrak{g} = \mathfrak{su}(N)$, comparing the resulting brane picture with expectations from the dimension formula.  At the end of section \ref{sec:suN} we point out an apparent mismatch between the dimension formula and the brane picture for the generic \tHooft defect.  This discrepancy is resolved in section \ref{sec:stuck} by considering the BIon description of \tHooft defects, where we also describe the processes of monopole extraction and bubbling.  Finally in section \ref{sec:branejump} we discuss the wall-crossing properties of the index in terms of brane realizations.

\section{Singular monopoles and the dimension formula}\label{sec:review}

Singular monopoles on $\mathbb{R}^3$ are solutions to the Bogomolny equation, $F = \pm \star \eD \Phi$, on $\UU := \mathbb{R}^3 \setminus \{ \vec{x}_n \}_{n=1}^{N_t}$, satisfying specific boundary conditions as $\vec{x} \to \vec{x}_n$ and as $|\vec{x}| \to \infty$.  Here $F = \ed A + A \wedge A$ and $\eD \Phi = \ed \Phi + [A, \Phi]$.  $(A,\Phi)$ are a $\mathfrak{g}$-valued gauge field and scalar field, where $\mathfrak{g}$ is the Lie algebra of a simple compact gauge group $G$.  We work in geometric conventions where elements of $\mathfrak{g}$ are represented by anti-Hermitian matrices.  Further details on our Lie algebra and Lie group conventions can be found in appendix A of \cite{MRVdimP1}.  In the absence of \tHooft defects, one can take either sign in the Bogomolny equation.  The plus sign, say, corresponds to studying monopole configurations, while the minus corresponds to anti-monopoles.  Let us denote this sign by $\sigma$, so that the Bogomolny equation reads $F = \star \eD (\sigma \Phi)$.  

The insertion of an \tHooft defect at position $\vec{x}_n$ is defined by the specification of boundary conditions \cite{'tHooft:1977hy,Kapustin:2005py}.  The boundary conditions depend on three pieces of data: the position $\vec{x}_n$, the charge $P_n$, and a sign choice $\sigma$.  They are given by
\begin{equation}\label{tHooftpole}
\sigma \Phi = -P_{n} \frac{1}{2r_{n}} + O(r_{n}^{-1/2})~, \qquad F = \half P_n \sin{\theta}_{n} \ed \theta_{n} \ed \phi_{n} + O(r_{n}^{-3/2})~, \quad \textrm{as $r_n \to 0$}~,
\end{equation}
where $(r_{n} = |\vec{x} - \vec{x}_n|, \theta_{n},\phi_{n})$ are standard spherical coordinates centered on the defect.  Here, $P_n$ is a covariantly constant section of the adjoint bundle over the infinitesimal two-sphere surrounding $\vec{x}_n$.  However, by making patch-wise local gauge transformations, we can conjugate $P_n$ to a constant, valued in the Cartan sub-algebra $\mathfrak{t}$.  We will assume this has been done.  $P_n \in \mathfrak{t}$ then determines the transition function, $\mathpzc{g}$, on the overlap of the northern and southern patches of the infinitesimal two-sphere, $\mathpzc{g} = \exp(P_n \phi_n)$.  Single-valuedness of the transition function implies $\exp(2\pi P_n) = 1_{G}$, the identity element in $G$ and thus, by definition, $P_n$ resides in the co-character lattice $\Lambda_{G} \subset \mathfrak{t}$ of $G$.  We may think of the 't Hooft  defect as a Dirac monopole embedded into the gauge group $G$, where $P_{n}$ determines the embedding $U(1) \hookrightarrow T \subset G$ of $U(1)$ into a Cartan torus of $G$.  The sign $\sigma$ in \eqref{tHooftpole} must be the same as the one appearing in the Bogomolny equation, and if we have multiple defects we must choose the same sign for each in order for solutions to exist.

In \cite{MRVdimP1} we showed how to derive \eqref{tHooftpole} from a variational principle.  One takes the standard action for Yang--Mills--Higgs theory in the BPS limit of vanishing potential \cite{Bogomolny:1975de,Prasad:1975kr}, and adds certain boundary terms localized at the defects.  The boundary terms are chosen so as to render the variational principle well-defined, and this also provides an explanation of the allowed subleading behavior in \eqref{tHooftpole}.  As a bonus, the boundary terms in the action imply boundary terms in the Hamiltonian that give a regularized definition of the energy.  With this definition, singular monopole configurations will have finite energy under precisely the same asymptotic conditions that are imposed on smooth monopoles.  These conditions are
\begin{equation}\label{largerbc}
\Phi = \Phi_\infty - \frac{\sigma}{2r} \gm + O(r^{-(1 + \updelta)})~, \quad F = \half \gm \sin{\theta} \ed\theta \ed\phi + O(r^{-(2+\updelta)})~, \quad \textrm{as $r \to \infty$}~,
\end{equation}
for any $\updelta > 0$ and where $(r,\theta,\phi)$ are standard spherical coordinates on $\mathbbm{R}^3$.  Here $\Phi_\infty, \gm$ are covariantly constant, commuting sections of the adjoint bundle restricted to the two-sphere at infinity, but by making patch-wise (global) gauge transformations we may assume them to be constants valued in the Cartan subalgebra.

The magnetic charge $\gm$ determines a transition function $\mathpzc{g}_\infty = \exp(\gm \phi)$ for the $G$-bundle restricted to the two-sphere at infinity.  As with the case of the \tHooft charges, single-valuedness implies $\gm \in \Lambda_G$.  In the absence of \tHooft defects, the requirement that this transition function be extendable over all of $\mathbb{R}^3$ leads to the stronger quantization condition that $\gm$ is an element of the co-root lattice, $\gm \in \Lambda_{\rm cr}$.  This is in general a coarser lattice, and is equal to the co-character lattice when $G$ is simply-connected; we have $\Lambda_G/ \Lambda_{\rm cr} \cong \pi_1(G)$.  When \tHooft defects are present we conclude instead that the difference $\tilde{\gamma}_{\rm m} \equiv \gm - \sum_{n} P_n \in \Lambda_{\rm cr}$.

The moduli space $\fMM$ is the space of gauge equivalence classes of solutions to $F = \star \eD(\sigma \Phi)$ satisfying \eqref{tHooftpole} and \eqref{largerbc}:
\begin{align}\label{Mdefsigma}
& \fMM_{\sigma}\left( (\vec{x}_{n}, P_{n} )_{n = 1}^{N_t} ; \gamma_{\rm m} ; \Phi_{\infty} \right) := \cr
&~  \left\{ (A,\Phi) ~ \bigg| ~ F = \star \eD(\sigma \Phi)~, \begin{array}{l} \Phi = - \frac{\sigma }{2 |\vec{x} - \vec{x}_n|} P_n + O(|\vec{x}-\vec{x}_n|^{-1/2})~,~ \vec{x} \to \vec{x}_n~, \\  \Phi = \Phi_\infty - \frac{\sigma}{2|\vec{x}|} \gm + O(|\vec{x}|^{-(1 + \updelta)})~, ~ |\vec{x}| \to \infty \end{array}  \right\} \bigg/ \GG_{\{ P_n \}}~. \qquad ~ ~
\end{align}
A comment is due regarding the group of local gauge transformations, $\GG_{ \{P_n \} }$ .  As usual, these are gauge transformations that go to the identity at infinity.  Furthermore we take them to leave the \tHooft charges invariant.  If $\mathpzc{g} \in \GG_{ \{P_n\} }$ and $\mathpzc{g}_n$ is the restriction of $\mathpzc{g}$ to the infinitesimal two-sphere surrounding $\vec{x}_n$, then we require the adjoint action of $\mathpzc{g}_n$ on $P_n$ to leave $P_n$ fixed.  See \cite{MRVdimP1} for further details.  Although two \tHooft charges related by a Weyl transformation are physically equivalent, it is more convenient to work with fixed representatives than to define $\fMM$ in terms of Weyl orbits directly.  This is what we have done in \eqref{Mdefsigma}.

We have defined the space for either choice of sign $\sigma$, however one notices that the definition only depends on the combination $\sigma \Phi_\infty$.  This motivates defining a new Higgs field:
\begin{equation}\label{Xdef}
X := \sigma \Phi~,
\end{equation}
in terms of which the definition of the moduli space is
\begin{align}\label{Mdef}
& \fMM\left( (\vec{x}_{n}, P_{n} )_{n = 1}^{N_t} ; \gamma_{\rm m} ; X_{\infty} \right) := \cr
& ~  \left\{ (A,X) ~ \bigg| ~ F = \star \eD X~,\begin{array}{l} X = - \frac{1}{2 |\vec{x} - \vec{x}_n|} P_n + O(|\vec{x}-\vec{x}_n|^{-1/2})~,~ \vec{x} \to \vec{x}_n~, \\  X = X_\infty - \frac{1}{2|\vec{x}|} \gm + O(|\vec{x}|^{-(1 + \updelta)})~, ~ |\vec{x}| \to \infty \end{array}  \right\} \bigg/ \GG_{\{ P_n \}}~. \qquad ~ ~
\end{align}
We work mostly with the definitions \eqref{Xdef} and \eqref{Mdef} in the remainder of the paper.

In order to determine the dimension of $\fMM$, one studies the linear deformation problem for the Bogomolny equation.  For this purpose it is convenient to repackage the gauge and Higgs field into a four-component gauge field, $\hat{A}_a = (A_i, X)$.  We can define $\hat{A} = \hat{A}_a \ed x^a$ as a gauge field on a four-dimensional Euclidean space, $\UU \times S^1$, which is invariant under translation of the circle coordinate $x^4$.  We choose the orientation on this four-dimensional space such that $\ed^3 x \wedge \ed x^4$ is positive, and we take the metric to be flat, $\ed s^2 = \ed x_i \ed x^i + (\ed x^4)^2$.  Then the Bogomolny equation on $\UU$ is equivalent to the self-dual instanton equation on $\UU \times S^1$, $\hat{\star} \hat{F} = \hat{F}$.  If $\delta \hat{A}$ is an infinitesimal deformation, then $\hat{A}_a + \delta \hat{A}_a$ will be a solution when $\hat{A}_a$ is, provided the deformation satisfies the linearized equation
\begin{equation}\label{linearsd}
\hat{D}_{[a} \delta \hat{A}_{b]} = \half \epsilon_{ab}^{\phantom{ab}cd} \hat{D}_c \delta \hat{A}_d~,
\end{equation}
where $\hat{D}$ is the covariant derivative with respect to the background solution $\hat{A}$.  However, this deformation will only correspond to a tangent vector $\delta \in T_{[\hat{A}]} \fMM$, if it is not pure gauge.  In order to formulate this condition it is useful to introduce a metric on the space of finite-energy field configurations and require $\delta \hat{A}_a$ to be orthogonal to infinitesimal gauge transformations.  There is a natural metric that is induced from the flat metric on field configuration space, with respect to which the gauge orthogonality condition is
\begin{equation}\label{gaugeorth}
\hat{D}^a \delta \hat{A}_a = 0~.
\end{equation}
Then the dimension of $T_{[\hat{A}]} \fMM$ is given by the number of linearly independent square-normalizable modes $\delta \hat{A} \in \LL^2[\UU, \mathbb{R}^4 \otimes \mathfrak{g}]$ that are simultaneous solutions of \eqref{linearsd} and \eqref{gaugeorth}.

This comprises a set of four linearly independent equations that can be repackaged into a Dirac equation on $\UU$ as follows.  Introduce Euclidean sigma matrices $\tau^a = (\sigma^i, -i \mathbbm{1}_2)$ and $\bar{\tau}^a = (\sigma^i, i \mathbbm{1}_2)$, where the $\sigma^i$ are Pauli matrices.  Define the bi-spinor $\delta \hat{A}$ with components $\delta \hat{A}_{\alpha\dot{\beta}} := (\tau^a)_{\alpha\dot{\beta}} \delta \hat{A}_a$ and the Dirac operator $L :=  i \bar{\tau}^a \hat{D}_a$.  Then one can show that \eqref{linearsd} and \eqref{gaugeorth} hold if and only if $L \delta \hat{A} = 0$.  The adjoint of the operator $L$, acting on the Hilbert space $\LL^2[\UU, \mathbb{C}^2 \otimes \mathfrak{g} ]$, is $L^\dag = i \tau^a \hat{D}_a$.  For each $\psi \in \ker{L}$, one gets two solutions to \eqref{linearsd} and \eqref{gaugeorth} by taking either $\psi_\alpha = \delta \hat{A}_{\alpha\dot{1}}$ or $\psi_\alpha = \delta \hat{A}_{\alpha \dot{2}}$.  One can also show that $L L^\dag = -\hat{D}^2$, a positive-definite operator, and hence $\ker{L^\dag} =  \ker{L L^\dag} = 0$.  It follows that
\begin{align}\label{Dbarindex}
\dim T_{[\hat{A}]} \fMM =&~ 2 \dim{\ker{L}} = 2 \left( \dim{\ker{L}} - \dim{\ker{L^\dag}} \right) = 2 \ind{L}~.
\end{align}

In \cite{MRVdimP1} we recalled Weinberg's computation of the dimension in the case of smooth monopoles \cite{Weinberg:1979ma,Weinberg:1979zt} and generalized it to the case with \tHooft defects.  This uses techniques of Callias \cite{Callias:1977kg} to write the index of $L$ as the integral over $\UU$ of the divergence of a certain current, $J^i$, constructed from the Green's function of a closely related operator.  The index reduces to a sum of boundary terms, one for the asymptotic two-sphere and one each of the infinitesimal two-spheres surrounding the \tHooft defects.  Near each of the boundaries, using the limiting form of the Dirac operator, one can explicitly determine the leading behavior of the current $J^i$ that contributes to the index.

We generalized the computation further, by considering an arbitrary finite-dimensional representation $\rho$ of $G$ and the Dirac operator
\begin{equation}\label{Lrho}
L_{\rho} = i \sigma^i  \otimes \left( \pd_i + \rho(A_i) \right) - \mathbbm{1}_2 \otimes \rho(X)~.
\end{equation}
Taking $\rho$ to be the adjoint representation, we recover the operator that is relevant for the dimension of the moduli space, $L_{\rm ad} = L$, but $L_\rho$ is also a physically interesting operator in general.  If we couple fermions to the Yang--Mills--Higgs theory in a way that is consistent with $\NN = 2$ supersymmetry, $L_\rho$ is the operator that controls the spectrum of these fermions in the background of the (singular) monopole configuration.  In particular, $\LL^2$-normalizable zero-modes of $L_\rho$ play an important role in the semiclassical analysis of such theories \cite{Manton:1993aa,Sethi:1995zm,Cederwall:1995bc,Gauntlett:1995fu,Henningson:1995hj}.  One can again show that $\ker{L_{\rho}^\dag} = \{0 \}$ and thus the dimension of the kernel of $L_\rho$ is given by the index.  Geometrically, over each point $[\hat{A}] \in \fMM$, $\ker{L_\rho}$ is a finite-dimensional vector space.  These vector spaces patch together to form a vector bundle, which is precisely the index bundle of the operator $L_\rho$, \cite{Taubes:1984je,Manton:1993aa}.  For $\rho = \ad$, and in the $\NN = 2$ theory, the index bundle is isomorphic to the tangent bundle.\footnote{This is possible in the $\NN = 2$ theory because there is a doublet of fermions that satisfy a reality condition.}

To construct the current that appears in the index theorem, let
\begin{equation}
\Gamma^a = \left( \begin{array}{c c} 0 & \tau^a \\ \bar{\tau}^a & 0 \end{array} \right)~,
\end{equation}
and define
\begin{equation}\label{hatDirac}
i \hat{\slashed{D}}_\rho := i \Gamma^a \rho(\hat{D}_{a}) = \left( \begin{array}{c c} 0 & i \tau^a \rho(\hat{D}_{a}) \\ i \bar{\tau}^a \rho(\hat{D}_{a}) & 0 \end{array} \right) = \left( \begin{array}{c c} 0 & L_{\rho}^\dag \\ L_\rho & 0 \end{array} \right)~.
\end{equation}
$i\hat{\slashed{D}}_\rho$ is a self-adjoint operator on a dense domain of the Hilbert space $\LL^2[\UU,\mathbb{C}^4 \otimes V_\rho]$, where $V_\rho$ is the representation space for $\rho$, $\rho : G \to GL(V_\rho)$.  Let $G_{\lambda}(\vec{x},\vec{y})$ be the integral kernel for the resolvent (\ie\ Green's function),
\begin{equation}
G_\lambda := \left( (i \hat{\slashed{D}})_\rho + \lambda \right)^{-1}~.
\end{equation}
Then the current $J^i$ is
\begin{equation}
J_{z,\rho}^i(\vec{x},\vec{y}) := i \tr_{\mathbb{C}^4 \otimes V_\rho} \left\{ \bar{\Gamma} \Gamma^i G_{i\sqrt{z}}(\vec{x},\vec{y}) \right\}~,
\end{equation}
where $\bar{\Gamma} = \Gamma^1 \Gamma^2 \Gamma^3 \Gamma^4$, and the index theorem takes the form
\begin{align}\label{index1}
\ind{L_\rho} =&~ \lim_{z \to 0} I_\rho(z)~, \qquad \textrm{where} \cr
I_\rho(z) =&~ \half \left( \ \lim_{r = |\vec{x}|\to \infty} - \sum_{n=1}^{N_t} \ \lim_{r = |\vec{x} - \vec{x}_n| \to 0} \right) \int_{S^2} \vol_{S^2} r^2 \hat{r} \cdot \vec{J}_{\rho,z}(\vec{x},\vec{x})~.
\end{align}
The minus sign takes into account the relative orientation of the boundary components of $\UU$, $\vol_{S^2}$ is the volume form on the unit two-sphere, and we are using a spherical coordinate system centered on $\vec{x} = 0$ for the asymptotic sphere and $\vec{x} = \vec{x}_n$ for the infinitesimal ones.  The asymptotic term was evaluated by Weinberg \cite{Weinberg:1979ma,Weinberg:1979zt}, and in \cite{MRVdimP1} we evaluated the terms associated with defects.  The final result is expressed purely in terms of the data $\left( \{ P_n \} ; \gm; X_\infty \right)$.

Let $\Delta_\rho$ denote the set of weights of the representation $\rho$ and let $n_\rho(\mu)$ denote the degeneracy of the weight $\mu \in \Delta_\rho$.  Then our formula for the index is
\begin{align}
\ind{L_\rho} =&~ \lim_{z \to 0^+}  \half  \sum_{\mu \in \Delta_{\rho}} n_\rho(\mu) \left\{ \frac{ \langle \mu, X_\infty \rangle \langle \mu, \gm \rangle }{\sqrt{ \langle \mu, X_\infty \rangle^2 + z} } + \sum_{n=1}^{N_t} | \langle \mu, P_n \rangle | \right\}~.
\end{align}
This formula is derived under the assumption $\langle \mu, X_\infty \rangle \neq 0$ for all $\mu \neq 0$.  Thus we can write
\begin{align}\label{indLrho}
\ind{L_\rho} =&~ \half  \sum_{\mu \in \Delta_{\rho}} n_\rho(\mu) \left\{ \sgn\left( \langle \mu, X_\infty \rangle \right) \langle \mu, \gm \rangle  + \sum_{n=1}^{N_t} | \langle \mu, P_n \rangle | \right\}~,
\end{align}
where we understand the contribution of the zero weight to be zero.  The term involving the asymptotic data $(\gm; X_\infty)$ originates from the asymptotic two-sphere boundary term, while the term involving $P_n$ originates from the boundary term associated with the infinitesimal two-sphere surrounding $\vec{x}_n$.  Note that $\sum_{\mu \in \Delta_\rho} |\langle \mu, P_n \rangle |$ is the trace of the diagonal matrix $| \rho(P_n) |$ and is thus Weyl invariant.  This is consistent with the fact it is only the Weyl orbit of $P_n$ that is physical.  In particular we are free to replace $P_n$ with $P_{n}^-$ in the above expression.

Now consider the adjoint representation, where the nonzero weights are the roots.  $X_\infty$ defines a half-space that splits the roots into positive and negative, $\Delta = \Delta^+ \cup \Delta^-$, where $\alpha \in \Delta^+ \iff \langle \alpha, X_\infty \rangle > 0$.  Furthermore $\alpha \in \Delta^+ \iff - \alpha \in \Delta^-$ and $n(\alpha) = 1$ for all roots $\alpha$.  Under these circumstances the expression \eqref{indLrho} can be replaced with twice the sum over positive roots, in which case the $\sgn(\langle \alpha, X_\infty \rangle )$ factor can be dropped.  Thus for the dimension we find
\begin{align}\label{dim2}
\dim_{\mathbb{R}} \fMM =&~ 2 \ind{L}  = 2 \sum_{\alpha \in \Delta^+} \left\{ \langle \alpha, \gm \rangle + \sum_{n=1}^{N_t} | \langle \alpha, P_{n}^- \rangle | \right\} \cr
=&~2  \sum_{\alpha \in \Delta^+} \left\{ \langle \alpha, \gm \rangle - \sum_{n=1}^{N_t} \langle \alpha, P_{n}^- \rangle \right\}  = 2 \sum_{\alpha \in \Delta^+} \langle \alpha, \tilde{\gamma}_{\rm m} \rangle \cr
=&~ 4 \langle \varrho, \tilde{\gamma}_{\rm m} \rangle = 4 \sum_{I = 1}^{\rnk{\mathfrak{g}}} \tilde{m}^I ~.
\end{align}
Here we first replaced $P_n$ with $P_{n}^-$ in \eqref{indLrho} using Weyl invariance and then we noted that $|\langle \alpha, P_{n}^- \rangle | = - \langle \alpha, P_{n}^- \rangle$ for all positive roots $\alpha$.  Next we recalled the definition of the relative magnetic charge, and then the definition of the Weyl vector, $\varrho := \half \sum_{\Delta^+} \alpha$.  We expanded $\tilde{\gamma}_{\rm m} = \sum_I \tilde{m}^I H_I$ and used the property $\langle \varrho, H_I \rangle = 1$ for all simple co-roots $H_I$.
 
More generally, for an arbitrary representation $\rho$, we showed in \cite{MRVdimP1} that \eqref{indLrho} can be written in the form
\begin{equation}\label{nonnegP1}
\ind{L_\rho} = \sum_{ \mathclap{\substack{ \mu \in \Delta_\rho \\ \langle \mu, X_\infty \rangle > 0 }} } n_{\rho}(\mu) \langle \mu, \tilde{\gamma}_{\rm m} \rangle + \sum_{n=1}^{N_t} \half \sum_{\mu \in \Delta_\rho} n_{\rho}(\mu) \left( \langle \mu, P_{n}^- \rangle + |\langle \mu, P_{n}^- \rangle | \right)~,
\end{equation}
and this makes it manifest that the index is an integer.  In fact we know that \eqref{indLrho}, and hence \eqref{nonnegP1}, must be non-negative as well, since $\ker{L_{\rho}^\dag} = \{0\}$.

Finally, we note that \eqref{indLrho} possesses wall-crossing properties, as we vary the Higgs vev $X_\infty$ across walls where $\langle \mu, X_\infty \rangle = 0$ for some weight in the representation $\rho$.  Consider such a wall and let $\mu_a$
be the (parallel) weights which all define the same wall.
As $X_\infty$ crosses this wall some quantities
$\sgn(\langle \mu_a, X_\infty \rangle)$ change from $-1$ to $+1$
and some change from $+1$ to $-1$. Let $\chi_a=+1$ in the former case
and $\chi_a=-1$ in the latter case.  Then the difference in the index
after the wall minus before the wall is
\be\label{eq:IndexJump}
\Delta \ind L_\rho = \sum_{a} \chi_a n_\rho(\mu_a) \langle \mu_a, \gm \rangle.
\ee
This has some interesting physical and geometrical implications and interpretations.  If we are considering matter fermions in a representation $\rho$, this corresponds to some $\LL^2$ normalizable modes (\ie\ bound states of the fermions coupled to the monopole configuration) leaving or entering the spectrum.  Geometrically, the rank of the index bundle jumps.  In the case of the adjoint representation the rank of the tangent bundle jumps; in other words the moduli space itself changes.

This concludes our review of \cite{MRVdimP1}.  Now we are ready to discuss D-brane realizations of (singular) monopole configurations, which provide physical insight into the index and dimension formulae.  In addition we will use them to provide a concrete physical realization of the wall-crossing phenomena just discussed.

\section{Monopoles from branes}\label{sec:branereview}

First let us recall the story for smooth monopole configurations.  A more extensive review of this material can be found in the final chapter of \cite{Weinberg:2006rq}.  The simplest and most natural way to engineer a four-dimensional Yang--Mills--Higgs system in string theory is on the worldvolume of D3-branes.  The actual low energy gauge theory on $N$ coincident D3-branes is maximally supersymmetric $\NN = 4$ super Yang--Mills theory with gauge group $U(N)$.  The field content consists of the gauge field, six adjoint-valued scalars, and four adjoint-valued Weyl fermions.  We take the D3-branes to span spacetime directions $x^\mu$, $\mu = 0,\ldots,3$.  When $N = 1$ we have an Abelian theory and the six scalar fields, $\Phi^m$, $m = 4,\ldots,9$, describe the profile of the D3-brane in the six remaining directions of the ten-dimensional spacetime, transverse to the brane.  Working classically and on-shell, when $N > 1$ it is the $N$ components of each $\Phi^m$ along the Cartan subalgebra that describe the profiles of the $N$ D3-branes.

Classically we can truncate the $\NN = 4$ theory to the Yang--Mills--Higgs system by restricting consideration to field configurations where the fermions and all but one of the six scalars are set to zero.  We let $\Phi^{m=4}$ be the remaining scalar and we set $\Phi := \Phi^4$.  This is a consistent truncation because there are no source terms in the equations of motion for the remaining fields that are built purely out of $(A_\mu,\Phi)$.  The vev $i \Phi_\infty \in i \cdot \mathfrak{t}_{\mathfrak{u}(N)} \cong \mathbb{R}^N$ determines the asymptotic position of the branes in the $x^4$ direction.\footnote{Here we are choosing the origin of the $x^4$ coordinate such that all D3-branes are at $x^4 = 0$ when $\Phi_\infty = 0$.  The D3-branes sit at a point in the remaining $x^5$-$x^9$ directions which we can take to be the origin.  Setting the remaining Higgs fields of the $\NN = 4$ theory to zero means we do not consider fluctuations in these directions.}  As Lie algebras, $\mathfrak{u}(N) \cong \mathfrak{u}(1) \oplus \mathfrak{su}(N)$; the physical interpretation of the $\mathfrak{u}(1)$ component of $\Phi$ is that it represents the position of the center of mass of the brane system along $x^4$.  At the level of the Lagrangian the center of mass factor decouples and we can focus on the $\mathfrak{su}(N)$ theory only.  However at the level of Lie groups, $\pi : U(1) \times SU(N) \to U(N)$ is an $N$-fold covering, and this will be important when we consider magnetic charge lattices below.

Turning to the description of monopoles, consider the simplest case of the Prasad--Sommerfield solution \cite{Prasad:1975kr} for a single monopole in $\mathfrak{su}(2)$ gauge theory.  Let $H$ denote the simple co-root and $E_{\pm}$ the raising and lowering operators, such that $[i H, E_\pm] = \pm 2 E_\pm$.  We take the vev to be $X_\infty = \frac{m_W}{2} H$, where $m_W > 0$  denotes the $W$-boson mass.  The solution is
\begin{align}\label{PSsoln}
X^{\rm PS} =&~ \left( \frac{m_{W}}{2} \coth{(m_W r)} - \frac{1}{2 r} \right) H~, \cr
A^{\rm PS} =&~ \AA H + W (-i E_+) + \Wbar (-i E_-)~, \qquad \textrm{with} \cr
\AA =&~ \half ( \pm 1 - \cos{\theta})\ed\phi~, \qquad W =  \frac{m_{W} r}{2 \sinh{(m_W r)}} e^{\pm i \phi} \left( - i \ed\theta + \sin{\theta} \ed\phi \right)~,
\end{align}
in terms of spherical coordinates with $r = |\vec{x} - \vec{x}_0|$, and where the $\pm$ refer to the northern/southern patch.  The position $\vec{x}_0$ of the monopole gives three moduli; the fourth can be generated by acting with a global gauge transformation that preserves the asymptotic Higgs field.  Recall that $X = \sigma \Phi$ and the sign choice $\sigma$ is related to whether we consider monopoles (solutions to $F = \star \eD \Phi$) or anti-monopoles (solutions to $F = - \star \eD \Phi$).

\begin{figure}
\begin{center}
\includegraphics{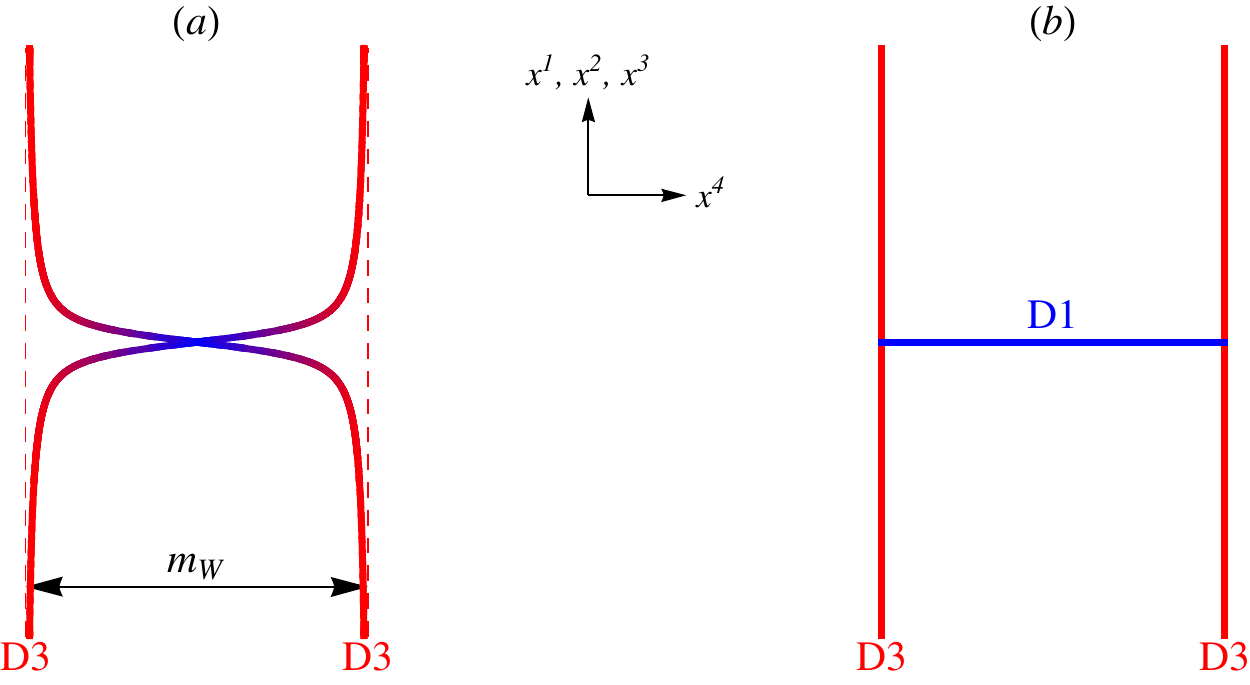}
\caption{Smooth $SU(2)$ mononpole and D1-string idealization.}
\label{fig1}
\end{center}
\end{figure}

The diagonal entries of the two-by-two matrix representation of $\Phi^{\rm PS} = \sigma X^{\rm PS}$ describe the profile of two D3-branes that bend towards each other and touch, as shown on the left of Figure \ref{fig1}.  As we dial up $m_W$ the bending becomes sharper and approaches the idealized picture on the right of Figure \ref{fig1}.  This picture suggests an alternative way of thinking about the monopole, as a D1-string stretched between the D3-branes.  Evidence in support of this picture was provided in \cite{Strominger:1995ac,Green:1996qg,Douglas:1996du}, where it was shown that D1-strings can end on D3-branes, depositing the appropriate unit of magnetic charge in the D3-brane worldvolume theory.  This point of view was greatly bolstered by Diaconescu \cite{Diaconescu:1996rk}, who showed that the BPS equations of the D1-string degrees of freedom are the Nahm equations describing the corresponding monopole in the ADHM--N construction.  Deriving the boundary conditions and jumping data of the Nahm construction from D-branes is subtle and has been discussed in several works \cite{Diaconescu:1996rk,Kapustin:1998pb,Tsimpis:1998zh,Chen:2002vb}.  Extensions of the construction to $SO$ and $Sp$ gauge groups making use of orientifold planes have also been considered \cite{Elitzur:1998ju,Ahn:1998ku}.

In the following we will simply use the D1-D3 system as a device for gaining intuition about monopoles and their moduli.  For example, motion on the four-dimensional moduli space of the  $SU(2)$ monopole is easy to visualize.\footnote{Note that while the gauge theory on the two D3-branes is a $U(2)$ theory, the monopole field configuration will be valued in $\mathfrak{su}(2) \subset \mathfrak{u}(2)$ provided we choose the origin of the $x^4$ direction to coincide with the center of mass position of the D3-branes, and that there is no net charge being deposited on them.}  It is well known that this moduli space is $\mathbb{R}^3 \times S^1$.  The $\mathbb{R}^3$ factor corresponds to moving the monopole in the three spatial directions; in the brane picture we are sliding the D1-string along the D3-branes.  The fourth modulus corresponds to the  global gauge transformation along the $U(1) \subset SU(2)$ preserving the Higgs field.  An electric field is generated in the D3-brane worldvolume theory when this modulus becomes time-dependent.  From the D1-string point of view this corresponds to exciting fundamental string states along the D1 to produce a dyonic D1-F1 string.  This is consistent with the fact that the endpoint of a fundamental string in the D3-brane acts as a source of electric charge.

Changing the sign of $\sigma$ corresponds to interchanging the D3-branes in Figure \ref{fig1}-$(a)$, and reversing the orientation of the D1-string in Figure \ref{fig1}-$(b)$.  A D1-string with opposite orientation is an anti-D1; if D1-strings represent monopoles then anti-D1-strings represent anti-monopoles.  We could denote this orientation choice by associating an arrow with each D1-string, but we will not do this since we will only consider brane configurations where all D1-strings have the same orientation (except in section \ref{sec:branejump}).  We will declare that the D1 corresponds to $\sigma = +$, so that $X = \Phi$ for all D-brane configurations in the following.

We can construct monopole configurations in a higher rank gauge group, $G$, by embedding the Prasad--Sommerfield solution along a root \cite{Weinberg:1979zt}.  Let $\varrho : SU(2) \hookrightarrow G$ such that
\begin{equation}\label{su2embed}
\varrho_{\ast}(H) = H_{\alpha}~, \qquad \varrho_{\ast}(E_{\pm}) = E_{\pm \alpha}~,
\end{equation}
for some fixed $\alpha \in \Delta^+ \subset \mathfrak{g}^\ast$.  The basic idea is to take $A = \varrho_\ast (A^{\rm PS})$ and $X = \varrho_{\ast}(X^{\rm PS}) + \Delta X_\infty$, where $\Delta X_\infty \in \mathfrak{t}$ is a constant ``correction factor'' so that $X$ asymptotes to some specified $X_\infty$ as $r \to \infty$.  To determine $m_{W}$ and $\Delta X_{\infty}$ note that $\Delta X_\infty$ must commute with $\varrho_\ast(A^{\rm PS})$ in order that the Bogomolny equation is satisfied.  This leads to a nontrivial condition coming from the $W$-bosons that
\begin{equation}
[E_{\pm \alpha} , \Delta X_\infty] =  \pm i \langle \alpha , \Delta X_\infty \rangle E_{\pm \alpha} = 0~.
\end{equation}
Therefore we take
\begin{equation}\label{DeltaX}
\Delta X_{\infty} = X_{\infty} - \frac{ \langle \alpha, X_\infty \rangle}{\langle \alpha, H_\alpha \rangle } H_\alpha = X_\infty - \half \langle \alpha, X_\infty \rangle H_\alpha~,
\end{equation}
Now we must choose $m_{W}$ so that $\varrho_{\ast}(X^{\rm PS})$ makes up the difference.  Since $\lim_{r \to \infty} \varrho_\ast(X^{\rm PS}) = \frac{m_{W}}{2} H_\alpha$, this means we should take $m_{W} = \langle \alpha, X_{\infty} \rangle$, which is indeed the mass of the $W$-boson along root $\alpha$.  In summary we have the following $G$-monopole solution:
\begin{align}\label{PSembed}
X =&~ \left( \half \langle \alpha, X_\infty \rangle \coth( \langle \alpha, X_\infty \rangle r ) - \frac{1}{2 r} \right) H_\alpha + X_\infty - \half \langle \alpha , X_{\infty} \rangle H_\alpha~, \cr
A =&~ \AA H_\alpha + W (-i E_\alpha) + \Wbar (-i E_{-\alpha})~, \qquad \textrm{with} \cr
\AA =&~ \half ( \pm 1 - \cos{\theta}) \ed\phi~, \qquad W = \frac{ \langle \alpha, X_{\infty} \rangle r }{2 \sinh(\langle \alpha, X_{\infty} \rangle r)} e^{\pm i \phi} ( - i \ed\theta + \sin{\theta} \ed\phi )~,
\end{align}
with asymptotic Higgs field $X_{\infty}$ and magnetic charge $\gm = H_\alpha$.

\begin{figure}
\begin{center}
\includegraphics{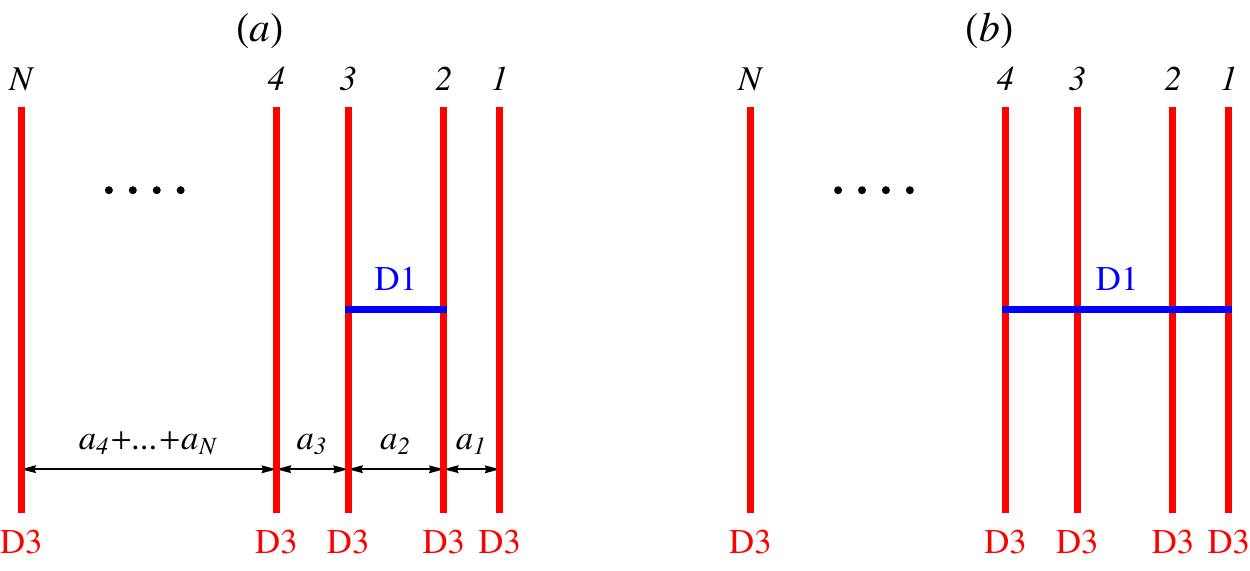}
\caption{(a) Embedding the Prasad--Sommerfield solution along the simple root $\alpha_2$.  (b) Embedding along the (non-simple) root $\alpha_{13} = \alpha_1 + \alpha_2 + \alpha_3$.  $a_I$ denotes the distance between the $I^{\rm th}$ and $(I+1)^{\rm th}$ D3-brane.}
\label{fig2}
\end{center}
\end{figure}

If $\alpha$ is a simple root then the dimension of the moduli space is four, while if $\alpha$ is non-simple this solution represents a spherically symmetric locus in a higher-dimensional moduli space corresponding to several fundamental monopoles sitting atop each other.  The D-brane picture of these two situations for gauge group $G = SU(N)$ is extremely intuitive and is described in Figure \ref{fig2}.  Note that positive roots for $SU(N)$ are in one-to-one correspondence with strings of simple roots, $\alpha_{IJ} := \alpha_I + \alpha_{I+1} + \cdots + \alpha_J$, for $1 \leq I \leq J \leq N-1$.  The cases $I = J$ correspond to the simple roots.  The fact that the moduli space is higher-dimensional in the non-simple case is obvious: the D1-string can break into smaller segments stretching between consecutive D3-branes which can move independently and carry their own electric charge.

\begin{figure}
\begin{center}
\includegraphics{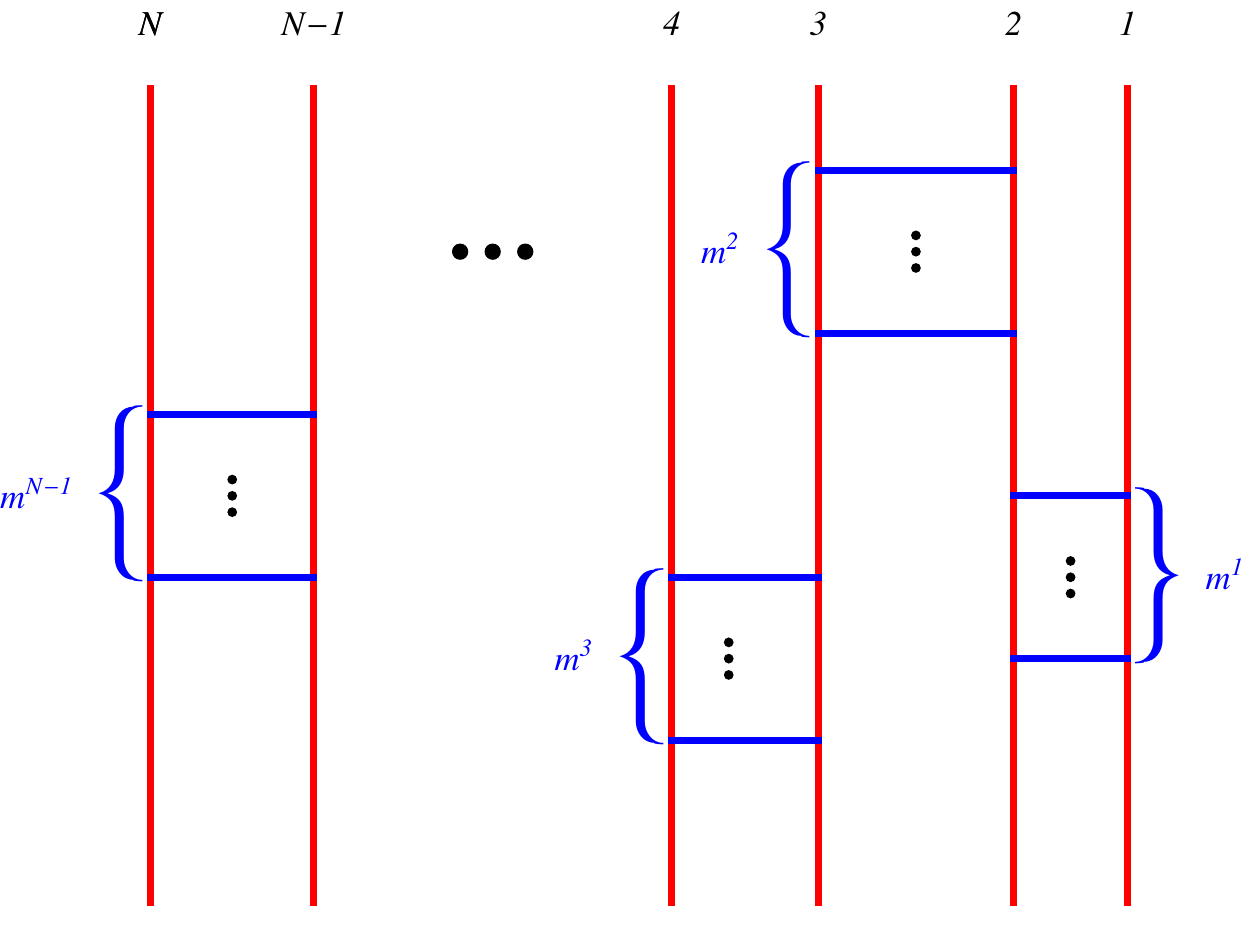}
\caption{A generic smooth monopole configuration in $SU(N)$ gauge theory with magnetic charge $\gm = \sum_{I=1}^{N-1} m^I H_I$.}
\label{fig3}
\end{center}
\end{figure}

Although we cannot write explicit field configurations, it is easy to draw an intersecting brane configuration describing a generic point on the moduli space of $G = SU(N)$ Yang--Mills--Higgs theory with specified asymptotic data $(\gm;X_\infty)$.  See Figure \ref{fig3}.  Requiring $X_\infty$ to be in the fundamental Weyl chamber means that we have an ordering of the D3-branes going from right (largest $x^4$-value) to left (smallest $x^4$-value).  The distance\footnote{Measured in units of the string length.} between the $I^{\rm th}$ and $(I+1)^{\rm th}$ brane is given by $a_I := \langle \alpha_I, X_\infty \rangle$.  There are $m^I$ D1-strings stretching between these branes, where $\gm = \sum_{I=1}^{N-1} m^I H_I$.  The condition that $m^I \geq 0$, $\forall I$, in order for solutions to the Bogomolny equation to exist is also natural: the D1-strings carry an orientation and all D1-strings should have the same orientation in order for the brane configuration to be BPS.

\section{Singular monopoles as limits of smooth ones}\label{sec:sm2si}

Singular monopoles---\ie\ configurations involving \tHooft defects---also have a natural realization via branes.  Consider a $U(N+1)$ configuration with a D1-string stretched from the $I^{\rm th}$ to the $(N+1)^{\rm th}$ brane.  If we send the $x^4$-position of the leftmost D3-brane to minus infinity, the D1-string becomes semi-infinite.  It was first pointed out in \cite{Hanany:1996ie} (in a T-dual context) that this should produce a singular monopole configuration.  This motivated the construction of some singular $SU(2)$ monopole moduli spaces in \cite{Cherkis:1997aa}, where the authors considered Nahm data for an $SU(3)$ theory with one of the intervals being semi-infinite.  Intuitively speaking, as the leftmost brane is sent off to infinity the corresponding $W$-bosons get infinitely heavy and what's left, given the form of \eqref{PSembed}, is the Dirac monopole gauge field and singular Higgs field of an \tHooft defect.  In addition to taking the limit there clearly needs to be some sort of projection to a gauge group whose rank is reduced by one.

Using the brane picture as a guide, our goal in this section is to give a precise prescription for constructing singular monopole configurations from smooth ones.  A subtlety when defects are present that we will need to address is how to ``factor out'' the center-of-mass $U(1)$ to obtain a singular monopole configuration for a simple gauge group $G$ with Lie algebra $\mathfrak{g} = \mathfrak{su}(N)$.

\begin{figure}
\begin{center}
\includegraphics{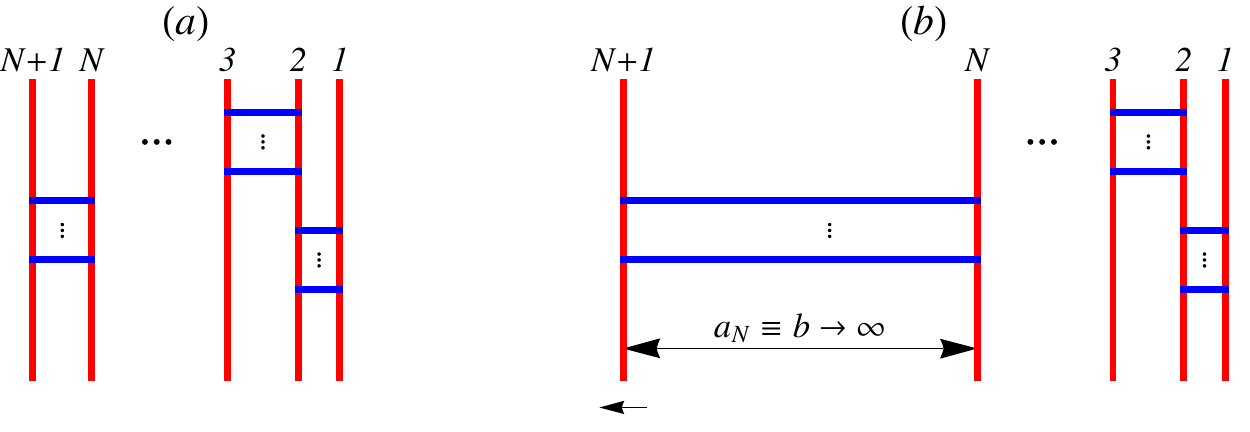}
\caption{$(a)$: A smooth monopole configuration in $\mathfrak{su}(N+1)$ gauge theory.  $(b)$: sending the leftmost D3-brane to $x^4 = -\infty$ to produce a singular monopole configuration in $\mathfrak{su}(N)$ gauge theory.}
\label{fig4}
\end{center}
\end{figure}

Let us begin by thinking about the brane motion depicted in Figure \ref{fig4} which corresponds to strongly breaking $\mathfrak{u}(N+1) \to \mathfrak{u}(N) \oplus \mathfrak{u}(1)$.  We are not interested in the decoupled $\mathfrak{u}(1)$ factor corresponding to the gauge and Higgs field on the brane being sent to infinity, so we project it out.  Representing the fields as anti-Hermitian matrices, this corresponds to the projection $\CC : \mathfrak{u}(N+1) \to \mathfrak{u}(N)$ given by
\begin{equation}
U_{N+1} = \left( \begin{array}{c c c | r} ~ & ~ & ~ & ~ \\ ~ & U_N & ~  & v \\  ~ & ~ & ~ & ~ \\ \hline ~ & -v^\dag & ~ & w \end{array} \right) \xmapsto{~~\CC~~} U_N~,
\end{equation}
where $U_N$ is the $N \times N$ upper left block of the $\mathfrak{u}(N+1)$ matrix $U_{N+1}$.  We would like to relate this to a projection $\mathfrak{su}(N+1) \to \mathfrak{su}(N)$.  However what we have canonically is an isomorphism $\mathfrak{u}(N) \cong \mathfrak{u}(1) \oplus \mathfrak{su}(N)$, defined by
\begin{align}\label{Phicon}
& \varPhi : \mathfrak{u}(N) \xrightarrow{\cong} \mathfrak{u}(1) \oplus \mathfrak{su}(N)~, \cr
& \varPhi(U) = \left( \frac{ \tr_{\bf N}(U) }{N}~,~ U - \frac{\tr_{\bf N}(U) }{N} \cdot \mathbbm{1}_N \right)~,
\end{align}
where $\tr_{\bf N}$ denotes the trace in the fundamental representation.  Thus we can construct a projection $\Pi' : \mathfrak{u}(1) \oplus \mathfrak{su}(N+1) \to \mathfrak{u}(1) \oplus \mathfrak{su}(N)$ by requiring that the following diagram commute:
\begin{equation}\label{Pitilde}
\xymatrix{
\mathfrak{u}(N+1)   \ar@{->}[rr]^{\CC}  \ar@{->}[d]_{\varPhi} & &  \mathfrak{u}(N) \ar@{->}[d]^{\varPhi}  \\   \mathfrak{u}(1) \oplus \mathfrak{su}(N+1) \ar@{-->}[rr]^{\Pi'}  & & \mathfrak{u}(1) \oplus \mathfrak{su}(N)~. }
\end{equation}
%

We can always start in the lower left corner of \eqref{Pitilde} with an element that has trivial $\mathfrak{u}(1)$ component, $(0,V_{N+1}) \in \mathfrak{u}(1) \oplus \mathfrak{su}(N+1)$.  Mapping to $\mathfrak{u}(n+1)$ with $\varPhi^{-1}$, this corresponds in the brane picture to choosing the center of mass position of the $N+1$ D3-branes to be at $x^4 = 0$, and starting in Figure \ref{fig4}-$(a)$ with no semi-infinite D1-strings.  Applying $\CC$ and then $\varPhi$ we find
\begin{equation}
(0,V_{N+1}) \xmapsto{~~\Pi'~~} \left( \frac{\tr_{\bf N}(V_N)}{N} ~,~ V_N - \frac{\tr_{\bf N}(V_N) }{N} \cdot \mathbbm{1}_N \right)~,
\end{equation}
where $V_N$ is the upper-left $N\times N$ block of $V_{N+1}$.  Note that, although we started with a configuration that had trivial $\mathfrak{u}(1)$ component in $\mathfrak{u}(N+1)$, we will generally have a nontrivial $\mathfrak{u}(1)$ component in the reduced theory.  In terms of branes, after pulling the leftmost brane to $-\infty$, the remaining branes generally will carry a net charge.  This charge is due to the resulting semi-infinite $D1$-strings.  There will be no net charge if there were no D1-strings attached to the leftmost D3-brane to begin with, such that the net charge on the subsystem of the other $N$ D3-branes was zero.

At the level of local field configurations---\ie\ Lie algebras---there is no problem with simply projecting out the $\mathfrak{u}(1)$; let us denote this projection $p : \mathfrak{u}(1) \oplus \mathfrak{su}(N) \to \mathfrak{su}(N)$.  Thus we can construct our desired projection map $\Pi : \mathfrak{su}(N+1) \to \mathfrak{su}(N)$, via $\Pi = p \circ \Pi' \circ \iota$, where $\iota : \mathfrak{su}(N+1) \hookrightarrow \mathfrak{u}(1) \oplus \mathfrak{su}(N+1)$ is the inclusion map.  It acts according to
\begin{equation}\label{Pidef1}
V_{N+1} = \left( \begin{array}{c c c | c} ~ & ~ & ~ & ~ \\ ~ & V_N & ~  & v \\  ~ & ~ & ~ & ~ \\ \hline ~ & -v^\dag & ~ & - \tr_{\bf N}(V_N) \end{array} \right) \xmapsto{~~\Pi~~}  V_N - \frac{\tr_{\bf N}(V_N) }{N} \cdot \mathbbm{1}_N~.
\end{equation}
An equivalent characterization of this map is as follows.  Given a root decomposition of $\mathfrak{su}(N+1)$ in terms of raising and lowering operators $E_{\pm \alpha}$, let $\Delta_{\rm heavy}$ denote the set of those roots that have $\alpha_N$ in their expansion in the basis of simple roots:
\begin{equation}
\Delta_{\rm heavy} = \left\{ \alpha \in \Delta ~\bigg|~ \alpha = \sum_{I} n^I \alpha_I \textrm{~with~} n_N > 0 \right\} \subset \Delta~.
\end{equation}
Furthermore let $h^{N} \in \mathfrak{t}$ denote the magnetic weight dual to $\alpha_N$, such that $\langle \alpha_I, h^{N} \rangle = {\delta_I}^N$.  Then we define the projection map by its action on the $E_\alpha$ and its action on the Cartan subalgebra:
\begin{align}\label{Pidef2}
\Pi(E_\alpha) =&~ \left\{ \begin{array}{l l} E_{\alpha}~, & \alpha \notin \Delta_{\rm heavy} \\ 0~, & \alpha \in \Delta_{\rm heavy}~ \end{array} \right. ~; \qquad \Pi(H) = H - \frac{(h^{N}, H )}{(h^{N}, h^{N})} h^{N}~, ~ \forall H \in \mathfrak{t}~.
\end{align}

It is straightforward to check that the definitions \eqref{Pidef1} and \eqref{Pidef2} are equivalent for $\mathfrak{su}(N+1) \to \mathfrak{su}(N)$.  The second definition is perhaps more physical.  In particular the projection onto the space orthogonal to $h^{N}$ in $\mathfrak{t}$ may be viewed as a projection onto the space orthogonal to the component of the asymptotic Higgs field that is becoming infinite.  This is because we can always write $X_\infty = \sum_I \langle \alpha_I, X_\infty \rangle h^{I}$, and we recall that the $\langle \alpha_I, X_\infty \rangle$ have the physical interpretation as the distance between the $I^{\rm th}$ and $(I+1)^{\rm th}$ D3-brane.  The second definition also clearly generalizes to any compact simple $\mathfrak{g}$, where $\alpha_N \to \alpha_{\rm heavy} \in \Delta$, some fixed root along which the Higgs vev will get an infinitely large component.  In this case we denote the projection $\Pi : \mathfrak{g} \to \check{\mathfrak{g}}$.  We generally use a `` $\check{}$ '' to distinguish quantities in the reduced theory.

Note that the map $\Pi$ is not a Lie algebra homomorphism: $\Pi([V,V']) \neq [\Pi(V), \Pi(V')]$.  This is easiest to see from \eqref{Pidef1}.  $\Pi([V,V'])$ will have pieces originating from $v v^\dag$ that are not present in $[\Pi(V), \Pi(V')]$.  This is not necessarily a problem because the projection is only part of the operation that we must perform to construct a singular monopole configuration from a smooth one.  We additionally have to take the limit $\langle \alpha_N, X_\infty \rangle \to -\infty$.  For the field configurations $(A,X)$ we are applying the map to, this is equivalent to sending $v \to 0$, since $v$ represents the $W$-bosons that are becoming infinitely massive.  Hence in this case the combined projection and limiting procedure does produce a Lie algebra homomorphism $\mathfrak{su}(N+1) \to \mathfrak{su}(N)$.  We will use this method to construct several examples of singular monopole configurations from smooth ones in the following sections.

First, however, we must discuss global issues associated with the analogous projection at the Lie group level.  This is important because both the 't Hooft charge and asymptotic charge of a singular monopole configuration sit in lattices that are sensitive to the global structure of the Lie group.  As we mentioned previously $U(1) \times SU(N)$ is a $\mathbb{Z}_N$ covering of $U(N)$.  The central extension
\begin{equation}
1 \to \mathbb{Z}_N \to U(1) \times SU(N) \xrightarrow{\pi} U(N) \to 1~,
\end{equation}
does not split: the $N$-fold covering is nontrivial and there is no way to ``factor out'' the $U(1)$ from $U(N)$ and recover $SU(N)$.  What we get instead is $PU(N) := U(N)/U(1)$, which is isomorphic to $PSU(N) := SU(N)/\mathbb{Z}_N$.  At the level of Lie algebras $\varPhi : \mathfrak{u}(1) \oplus \mathfrak{su}(N) \xrightarrow{\cong} \mathfrak{u}(N)$ exponentiates to a homomorphism $\hat{\varPhi} : U(1) \times SU(N) \to U(N)$. The inverse image of the $U(1)$ subgroup of $U(N)$ of scalar matrices is a nontrivial $N$-fold covering with elements in $SU(N)$.  ``Factoring out the $U(1)$'' corresponds to descending that homomorphism to an isomorphism $(U(1) \times SU(N))/U(1) \cong PSU(N) \cong U(N)/U(1)$. Although one can ``factor out'' the Lie algebra summand $\mathfrak{u}(1)$ in the field configurations this is not possible for the transition functions of the bundles, and hence the appropriate gauge group associated with singular monopole configurations constructed from the projection and limiting procedure described above is $\check{G} = PSU(N)$.

One can quickly see that this is indeed the case by considering a simple example.  Suppose we start with a configuration consisting of $N+1$ D3-branes and a single D1-string stretched between the two leftmost branes.  In terms of the $\mathfrak{su}(N+1)$ theory this configuration has asymptotic magnetic charge $\gm = H_N$.  We apply the projection map \eqref{Pidef2} using the normalization of the Killing form $(A,B) = - \tr_{\bf N+1}(AB)$ for $\mathfrak{g} = \mathfrak{su}(N+1)$.  Since $i h^{N} = (N+1)^{-1} \diag{(1,\ldots,1,-N)}$, we have $(h^{N},H_N) = 1$ and $(h^{N},h^{N}) = N/(N+1)$.  Thus
\begin{align}\label{PiHN}
\Pi(H_N) =&~ H_N - \frac{N+1}{N} h^{N} \cr
=&~ H_N - \frac{1}{N} \left( H_1 + 2 H_2 + \cdots N H_N \right) \cr
=&~ - \frac{1}{N} H_1 - \frac{2}{N} H_2 -  \cdots - \frac{N-1}{N} H_{N-1} \cr
=&~ - \check{h}^{N-1}~,
\end{align}
where in the last step we introduced the $(N-1)^{\rm th}$ magnetic weight for $\mathfrak{su}(N)$.  Thus $\Pi(H_N)$, which is both the \tHooft and asymptotic magnetic charge in this example, sits in the magnetic weight lattice of $\mathfrak{su}(N)$, which is the co-character lattice of the adjoint group, $\Pi(H_N) \in \Lambda_{PSU(N)}$.

It is well known that specifying a gauge group $G$ and the dynamical
field content does not fully define a Yang--Mills theory.
Rather, one must also specify the allowed defects in the theory. The
distinction between theories based on line defects was recently
emphasized in
\cite{Gaiotto:2010be, Aharony:2013hda}.  For the various theories based
on groups whose universal cover is $SU(N)$ one can define
the distinct theories using the brane pictures we are using in this
section. The distinction between the theories emerges from different
consistent ways of factoring out the $U(1)$ center of mass degree of
freedom. (These are based on different consistent sets of semi-infinite
F1-D1 dyonic strings defined by defining $z_{\rm m}$, the sum of the magnetic charges modulo $N$
together with $z_{\rm e}$, the sum of the electric charges modulo $N$, and choosing a
maximal Lagrangian subgroup of $\mathbb{Z}_N \oplus \mathbb{Z}_N$ with elements $(z_{\rm e},z_{\rm m})$.) It would be interesting to discuss the
semiclassical framed states in this general class of theories, but in
this paper we will limit our scope to the   pure \tHooft defects.   Hence the
relevant $PSU$ theory, in the notation of \cite{Aharony:2013hda}, is the
$(PSU)_{0}$ theory.

Now we are ready to apply the projection and limiting procedure discussed above to construct some singular monopole configurations from smooth ones.  The usefulness of this procedure is that it has a clear interpretation in terms of branes (Figure \ref{fig4}).  Thus we can read off from the picture what the dimension of the singular monopole moduli space should be, and compare this with what we get by applying the dimension formula.

\section{Smooth to singular for $SU(3) \to PSU(2)$}\label{sec:su3}

\begin{figure}
\begin{center}
\includegraphics{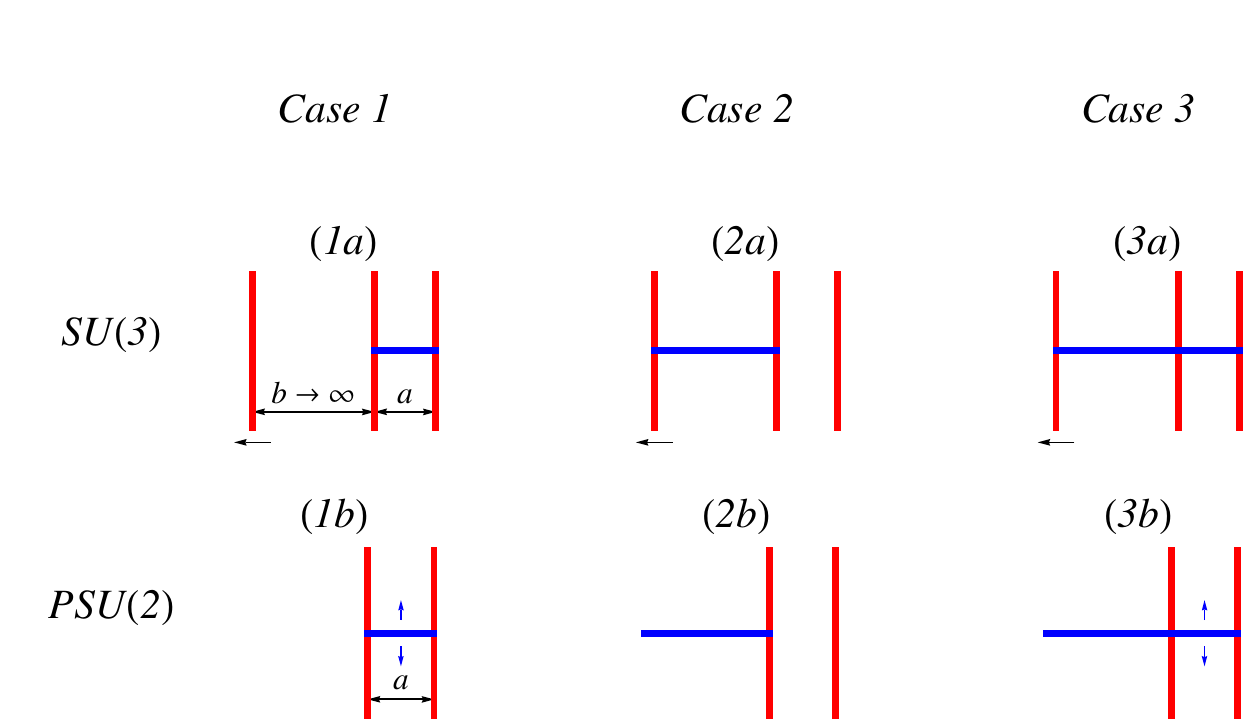}
\caption{Three cases of the smooth $\to$ singular limit for $SU(3) \to PSU(2)$.  In case 1 we obtain a smooth monopole in the reduced theory, in case 2 we obtain a pure \tHooft defect, and in case 3 we obtain a smooth monopole sitting on top of a pure \tHooft defect.}
\label{fig5}
\end{center}
\end{figure}

We begin with the simplest possible set of nontrivial examples, taking $\mathfrak{g} = \mathfrak{su}(3)$.  Let $\{ H_1, H_2 \}$ denote the basis of simple co-roots for the Cartan subalgebra.  There are raising and lowering operators for the corresponding simple roots, $E_{\pm \alpha_1}, E_{\pm \alpha_2}$, as well as raising and lowering operators for the non-simple root $\alpha_1 + \alpha_2$, $E_{\pm (\alpha_1 + \alpha_2)}$.  We have $\langle \alpha_1, H_1 \rangle = \langle \alpha_2, H_2 \rangle = 2$, while $\langle \alpha_1, H_2 \rangle = \langle \alpha_2, H_1\rangle = -1$.  Let $a,b \in \mathbb{R}_+$ parameterize the Higgs vev as follows:
\begin{align}\label{SU3vev}
X_{\infty} =&~ \frac{1}{3} (2 a + b) H_1 + \frac{1}{3} (a + 2 b) H_2 = a h^{1} + b h^{2} =  -\frac{i}{3} \left( \begin{array}{c c c} 2 a + b & 0 & 0 \\ 0 & -a + b & 0 \\  0 & 0 & - a -2b \end{array}\right)~,
\end{align}
where in the last step we expressed the vev in the fundamental representation.  We have
\begin{equation}
\langle \alpha_1, X_{\infty} \rangle = a~, \qquad  \langle \alpha_2, X_\infty \rangle = b~,
\end{equation}
such that $a$ corresponds to the distance between the center and right D3-brane and $b$ corresponds to the distance between the center and left D3-brane.  We will be taking the limit $b \to \infty$.  We consider each of the three cases in Figure \ref{fig5} in turn.

\subsection{case 1: $\gm = H_1$}

The first step is to write down the appropriate smooth monopole configuration corresponding to Figure \ref{fig5}-$(1a)$.  This requires \eqref{DeltaX} and \eqref{PSembed} with $\alpha = \alpha_1$ and $X_\infty$ given by \eqref{SU3vev}.  We find $\Delta X_\infty = \frac{1}{6} (a+2b) (H_1 + 2 H_2)$, which satisfies $\langle \alpha_1, \Delta X_\infty \rangle = 0$ as it should.  The monopole configuration is
\begin{align}\label{SU3smooth}
X =&~ \left( \frac{a}{2}  \coth(a r) - \frac{1}{2r} \right) H_1 + \frac{1}{6} (a + 2 b) (H_1 + 2 H_2 )~, \cr
A =&~ \AA H_{1} + W (-i E_{\alpha_{1}}) + \Wbar (-i E_{-\alpha_{1}})~, \qquad \textrm{with} \cr
\AA =&~ \half ( \pm 1 - \cos{\theta}) \ed\phi~, \qquad W = \frac{ a  r }{2 \sinh(a r)} e^{\pm i \phi} ( - i \ed\theta + \sin{\theta} \ed\phi )~.
\end{align}

The next step is the projection, \eqref{Pidef2}.  We have $(h^{2}, H_1) = 0$, $(h^{2}, H_2) = 1$, and $(h^{2},h^{2}) = 2/3$.  Thus acting on the co-roots we find
\begin{equation}\label{Picoroot}
\Pi(H_1) = H_1~, \qquad \Pi(H_2) = H_2 - \frac{3}{2} \cdot \frac{1}{3}(H_1 + 2 H_2) = - \frac{1}{2} H_1~.
\end{equation}
It follows that the projected fields in this case are
\begin{equation}
\Pi(A) = A~, \qquad \Pi(X) = \left( \frac{a}{2}  \coth(a r) - \frac{1}{2r} \right) H_1~.
\end{equation}
The final step, sending $b \to \infty$, is trivial in this case.  This field configuration, with $A$ given in \eqref{SU3smooth}, is simply the Prasad--Sommerfield solution, \eqref{PSsoln}.  In particular we get a smooth monopole configuration rather than a singular one.  This is consistent with the brane picture, Figure \ref{fig5}-$(1b)$, where there are no semi-infinite D1-strings.

\subsection{case 2: $\gm = H_2$}

The smooth monopole field configuration corresponding to Figure \ref{fig5}-$(2a)$ can be obtained from the previous one by interchanging $a \leftrightarrow b$ and the labels $1 \leftrightarrow 2$.  The result is
\begin{align}
X =&~ \left( \frac{b}{2}  \coth(b r) - \frac{1}{2 r} \right) H_2 + \frac{1}{6} (2a + b) (2 H_1 + H_2 )~, \cr
A =&~ \AA H_{2} + W (-i E_{\alpha_{2}}) + \Wbar (-i E_{-\alpha_{2}})~, \qquad \textrm{with} \cr
\AA =&~ \half ( \pm 1 - \cos{\theta}) \ed\phi~, \qquad W = \frac{ b  r }{2 \sinh(b r)} e^{\pm i \phi} ( - i \ed\theta + \sin{\theta} \ed\phi )~.
\end{align}
We apply the same projection map as before.  This time the W-bosons are eliminated.  Using \eqref{Picoroot} we find
\begin{align}
\Pi(X) =&~  \left( \frac{b}{4}  - \frac{b}{4} \coth(2 b r) + \frac{1}{4 r} \right) H_1 + \frac{a}{2} H_1~, \cr
\Pi(A) =&~   \AA \left(-\frac{1}{2} H_1\right) = - \frac{1}{4} (\pm 1 - \cos{\theta} \ed\phi) H_1~.
\end{align}

This configuration is not a solution to the Bogomolny equation; however, we have yet to take the limit $b \to \infty$.  Doing so, we find the solution
\begin{equation}\label{lineop1}
\lim_{b \to \infty} \Pi(X) = \frac{1}{4 r} H_1 + \frac{a}{2} H_1~, \qquad \lim_{b \to \infty} \Pi(A) = - \frac{1}{4} (\pm 1 - \cos{\theta} \ed\phi) H_1~.
\end{equation}
This is a $\check{\mathfrak{g}} = \mathfrak{su}(2)$ field configuration with 't Hooft defect boundary conditions at the origin.  The \tHooft charge and asymptotic magnetic charge are both given by
\begin{equation} \check{P} = - \half H_1 = \check{\gamma}_{\rm m}~.
\end{equation}
As expected, this charge is not in the co-character lattice of $SU(2)$ (which is equivalent to the co-root lattice), but it is in the co-character lattice of $PSU(2) \cong SO(3)$.

Let us apply the dimension formula \eqref{dim2}.  This is an example of a Cartan-valued solution where the asymptotic magnetic charge is in the anti-fundamental Weyl chamber: $\check{\gamma}_{\rm m} = \check{P} = \check{P}^-$.  Hence the relative charge is trivial, $\check{\tilde{\gamma}}_{\rm m} = 0$, and the dimension of the moduli space is zero.  This is consistent with the brane picture; there are no mobile D1-string segments present in Figure \ref{fig5}-$(2b)$.

\subsection{case 3: $\gm = H_1 + H_2$}

The moduli space of smooth monopoles in $\mathfrak{su}(3)$ gauge theory with magnetic charge $\gm = H_1 + H_2$ is eight-dimensional, and the metric is explicitly known \cite{Lee:1996kz}.  Two directions correspond to global $U(1) \times U(1) \subset SU(3)$ gauge transformations, while the remaining six directions are associated with the position of the two fundamental monopoles.  This is captured in the brane picture of Figure \ref{fig5}-$(3a)$ by the fact that there are two finite D1-string segments that can move and carry excited F1-string states independently.

As far as we are aware, the field configuration corresponding to a generic point on this moduli space has not been written down, although it could apparently be extracted from either the twistor analysis of \cite{Ward:1982ft,Athorne:1982ke}, or the ADHM--N analysis of \cite{Weinberg:1998hn}.  It would be interesting to do so and then apply our projection and limiting procedure to the result.  This should yield the explicit field configuration obtained in \cite{Cherkis:2007jm}, describing one smooth $\mathfrak{su}(2)$ monopole in the presence of an \tHooft defect.  Here we will content ourselves to study the special four-dimensional sublocus of spherically symmetric solutions obtained by embedding the Prasad--Sommerfield solution along the root $\alpha_1 + \alpha_2$.  In the brane picture this corresponds to aligning the two D1-string segments as depicted in Figure \ref{fig5}-$(3a)$.  For this special situation we find the embedded solution
\begin{align}
X =&~  \left( \frac{a+b}{2} \coth((a+b) r) - \frac{1}{2 r} \right) (H_1+H_2) + \frac{1}{6} (a-b) (H_1 - H_2 )~, \cr
A =&~ \AA (H_1+ H_{2}) + W (-i E_{\alpha_{1} + \alpha_2}) + \Wbar (-i E_{-(\alpha_{1} + \alpha_2)})~, \qquad \textrm{with} \cr
\AA =&~ \half ( \pm 1 - \cos{\theta}) \ed\phi~, \qquad W = \frac{ (a+b)  r }{2 \sinh((a+b) r)} e^{\pm i \phi} ( - i \ed\theta + \sin{\theta} \ed\phi )~.
\end{align}

Applying the projection and taking the $b \to \infty$ limit yields the $\check{\mathfrak{g}} = \mathfrak{su}(2)$  Cartan-valued solution
\begin{equation}\label{lineop2}
\lim_{b\to \infty} \Pi(X) = (a - \frac{1}{4 r}) H_1~, \qquad \lim_{b\to \infty} \Pi(A) = \frac{1}{4} (\pm 1 - \cos{\theta} \ed\phi) H_1~.
\end{equation}
The \tHooft and asymptotic magnetic charge,
\begin{equation}
\check{P} = \check{\gamma}_{\rm m} = \half H_1~,
\end{equation}
are again in the co-character lattice of $PSU(2)$.  In this case however $\check{P}$ is not in the anti-fundamental Weyl chamber and the relative charge is non-zero: $\check{\tilde{\gamma}}_{\rm m} = \check{\gamma}_{\rm m} - \check{P}^- = H_1$.  Note that $\check{\tilde{\gamma}}_{\rm m} \in \Lambda_{\rm cr}$ as expected. The dimension formula states that there should be a four-dimensional tangent space at this point in moduli space.  This is confirmed by the brane picture of Figure \ref{fig5}-$(3b)$, as there is one remaining finite-length D1-string segment.  The spherically symmetric solution we have written corresponds to the special point in moduli space where the finite length and semi-infinite D1-string segments are aligned.

It is worth noting that the \tHooft charges obtained in the previous two examples, $\check{P} = \pm \half H_1$, are related by a Weyl transformation and thus represent the same physical defect.  However two asymptotic charges related by a Weyl transformation are not physically equivalent (if we hold the Higgs vev fixed), and therefore the two configurations are physically distinct.

\section{Smooth to singular for $SU(N+1) \to PSU(N)$}\label{sec:suN}

Now that we have seen how things work in the simplest example, let us be slightly more general.  We will start with the system of $N+1$ D3-branes and pull the leftmost brane off to $x^4 = -\infty$, considering various choices for D1-string configurations.  Let $a_I = \langle \alpha_I, X_\infty \rangle \in \mathbb{R}_+$, $I = 1,\ldots,N$, denote the separation between the $I^{\rm th}$ and $(I+1)^{\rm th}$ brane.  We also denote the leftmost separation $b \equiv a_N$, which will be sent to infinity.

\begin{figure}
\begin{center}
\includegraphics{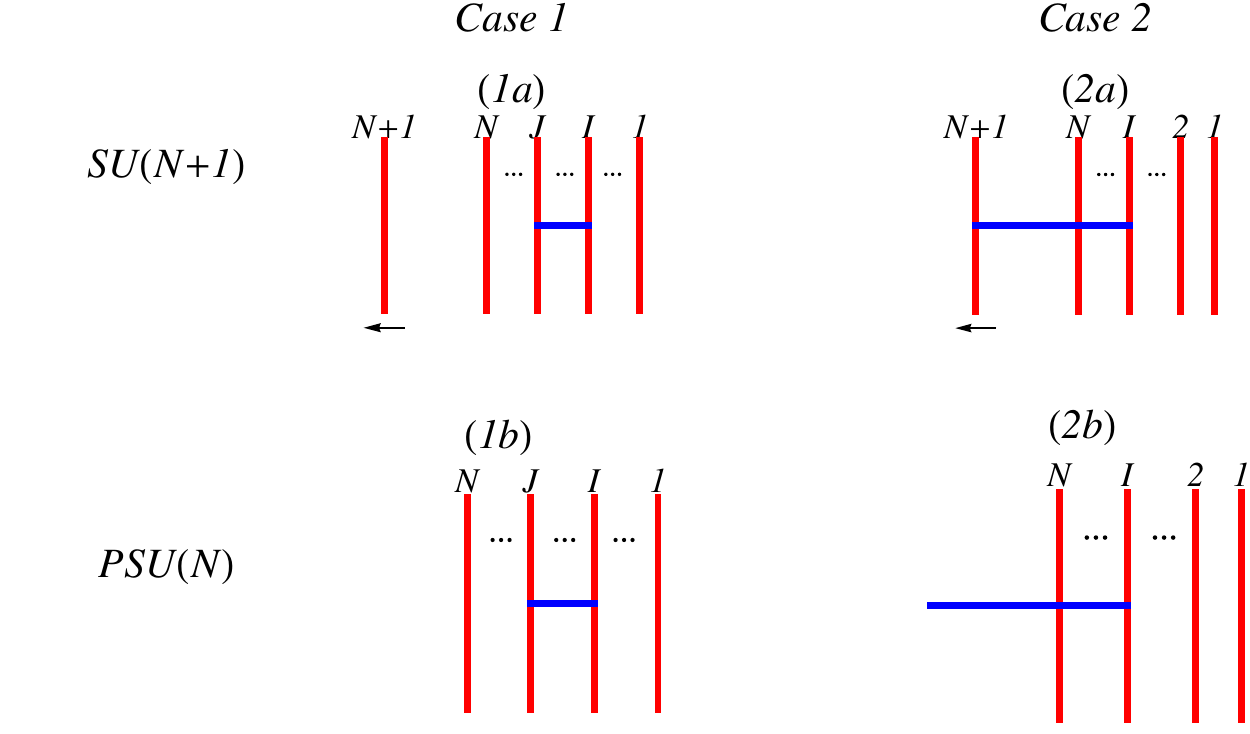}
\caption{Two cases of the smooth $\to$ singular limit for $SU(N+1) \to PSU(N)$.  In case 1 we obtain a smooth (in general composite, if $J > I+1$) monopole, while in case 2 we obtain a singular monopole configuration.}
\label{fig6}
\end{center}
\end{figure}

The components of the Cartan matrix for $\mathfrak{su}(N+1)$ are $C_{IJ} := \langle \alpha_I, H_J \rangle$.  Thus, denoting the components of the inverse of the Cartan matrix by $C^{IJ}$, the asymptotic Higgs vev is
\begin{equation}\label{SUNvev}
X_\infty = \sum_{I=1}^N a_I h^{I} = \sum_{I,J=1}^N C^{IJ} a_I H_J~.
\end{equation}
The Higgs field is acquiring an infinite component along the direction in $\mathfrak{t}$ spanned by the $N^{\rm th}$ fundamental magnetic weight,
\begin{equation}
h^{N} = \sum_{J=1}^N C^{NJ} H_J = \frac{1}{N+1} \left( H_1 + 2 H_2 + \cdots N H_N \right) = \frac{-i}{N+1}\left( \begin{array}{c c c c c} 1 & 0 & \cdots & & 0 \\ 0 & 1 & 0 & \cdots & 0 \\ \vdots & & \ddots & & \vdots \\ & & & 1 & 0 \\ 0 & \cdots & & 0 & -N \end{array} \right)~.
\end{equation}
We also recall that the full set of positive roots is given by $\Delta^+ = \{ \alpha_{IJ} := \sum_{K=I}^J \alpha_K~,~ \textrm{for} ~1 \leq I \leq J \leq N \}$, with $\alpha_{II} = \alpha_I$ the simple roots.  We denote the corresponding co-roots $H_{IJ} := H_{\alpha_{IJ}}$.  Indices $\check{I},\check{J}$ will run over the values $1,\ldots, N-1$.

We will consider two types of magnetic charge.  In the first case we take $\gm = H_{\check{I}\check{J}}$, a co-root that does not contain $H_N$ in its decomposition, corresponding to the brane configuration depicted in Figure \ref{fig6}-$(1a)$.  In this case we expect the projection and limiting procedure to produce a smooth monopole in the reduced theory.  In the second case we take $\gm = H_{I N}$, a co-root that does contain $H_N$, corresponding to the brane configuration depicted in Figure \ref{fig6}-$(2a)$.  Here we expect to obtain a singular monopole configuration.

In both of these cases we are considering magnetic charges that have all $m^I$ equal to zero or one.  Again the metric on moduli space is explicitly known for such configurations \cite{Lee:1996kz}, and although explicit field configurations representing a generic point on the moduli space have not been written down, it should be possible to obtain them via the Nahm transform \cite{Weinberg:1998hn}.  Our reason for choosing charges of this type is that the moduli space has a locus of spherically symmetric solutions where the field configurations are obtained by embedding the $SU(2)$ solution along the root.

In the last part we will discuss the most generic brane configuration describing a $PSU(N)$ singular monopole, where it is not possible to write explicit field configurations.  We will discuss how to appropriately assign \tHooft and asymptotic magnetic charges, and compare the quaternionic dimension of the moduli space computed from these charges with the number of mobile D1-string segments.

\subsection{case 1: $\gm = H_{\check{I} \check{J}}$}

This is the situation depicted in Figure \ref{fig6}-$(1)$.  The field configuration when all fundamental monopoles are coincident (all D1-string segments are aligned) is
\begin{align}
X =&~ \left( \frac{ a_{\check{I}\check{J}} }{2}  \coth( a_{\check{I} \check{J}} r) - \frac{1}{2 r} \right) H_{\check{I} \check{J}} + X_{\infty} - \frac{a_{\check{I}\check{J}}}{2} H_{\check{I}\check{J}}~, \qquad \cr
A =&~ \AA H_{\check{I}\check{J}} + W (-i E_{\alpha_{\check{I}\check{J}}}) + \Wbar (-i E_{-\alpha_{\check{I}\check{J}}})~, \qquad \textrm{where} \cr
\AA =&~  \half (\pm 1 - \cos{\theta})d\phi~, \qquad W = \frac{ a_{\check{I}\check{J}} r}{2 \sinh(a_{\check{I}\check{J}} r)} e^{\pm i\phi} (-i \ed\theta + \sin{\theta} \ed\phi)~,
\end{align}
where $a_{\check{I}\check{J}} := \langle \alpha_{\check{I}\check{J}}, X_\infty \rangle = a_{\check{I}} + \cdots + a_{\check{J}}$.
For the projection we note that $(h^{N}, H_{\check{I}}) = 0$, so $\Pi(H_{\check{I}\check{J}}) = H_{\check{I}\check{J}}$.  The $W$-bosons also survive since $\alpha_{\check{I}\check{J}} \notin \Delta_{\rm heavy}$.  Therefore the projection only acts nontrivially on the vev, $X_\infty$.  Using \eqref{SUNvev} and \eqref{PiHN} we have
\begin{align}
\Pi(X_\infty) =&~ \sum_{I,J=1}^N C^{IJ} a_I \Pi(H_J) = \sum_{I,J=1}^N C^{IJ} a_I H_J - \frac{N+1}{N} \sum_{I=1}^N C^{IN} a_I h^{N} \cr
=&~ \sum_{I,J = 1}^N \left( C^{IJ} - \frac{N+1}{N} C^{IN} C^{NJ} \right) a_I H_J \cr
=&~ \sum_{\check{I},\check{J} = 1}^{N-1} \left( C^{\check{I}\check{J}} - \frac{N+1}{N} C^{\check{I}N} C^{N\check{J}} \right) a_{\check{I}} H_{\check{J}} = \sum_{\check{I},\check{J} = 1}^{N-1} \check{C}^{\check{I}\check{J}} a_{\check{I}} H_{\check{J}} \equiv \check{X}_\infty~.
\end{align}
In going from the second to the third line we noted that, since $C^{NN} = N/(N+1)$, all terms with either $I = N$ or $J = N$ drop out.  We then observed that the quantity in parentheses is simply the inverse of the Cartan matrix for $SU(N)$.  The reduced Higgs vev, $\check{X}_\infty$, is in the fundamental Weyl chamber of $\check{\mathfrak{t}}$, the Cartan subalgebra of $\mathfrak{su}(N)$, since $a_{\check{I}} \in \mathbbm{R}_+$.  Hence the result of applying the projection is simply
\begin{align}\label{SUNsmooth}
\Pi(A) = A~, \qquad \Pi(X) = \left( \frac{ a_{\check{I}\check{J}} }{2}  \coth( a_{\check{I} \check{J}} r) - \frac{1}{2 r} \right) H_{\check{I} \check{J}} + \check{X}_{\infty} - \frac{a_{\check{I}\check{J}}}{2} H_{\check{I}\check{J}}~.
\end{align}
Note that all dependence on $b = a_N$ has dropped out so the limit is trivial.  The configuration \eqref{SUNsmooth} is precisely the smooth field configuration one gets by embedding the Prasad--Sommerfield solution into an $SU(N)$ gauge theory with asymptotic Higgs vev $\check{X}_\infty$ along the root $\check{\alpha}_{\check{I}\check{J}} = \alpha_{\check{I}\check{J}}$, and this is what one expects from the brane picture.

\subsection{case 2: $\gm = H_{IN}$}

This corresponds to the brane configuration depicted in Figure \ref{fig6}-$(2)$.  A special case is when $I=N$ so that $\gm = H_N$; this is the analog of $\gm = H_2$ in the previous section and was discussed around \eqref{PiHN}.

The field configuration for the spherically symmetric solution is
\begin{align}
X =&~ \left( \frac{ a_{I,N-1} + b}{2}  \coth\left((a_{I,N-1}+b)\right) - \frac{1}{2 r} \right) H_{IN} +  X_{\infty} - \frac{a_{I,N-1} + b}{2} H_{IN}~, \cr
A =&~ \AA H_{IN} + W (-i E_{\alpha_{IN}}) + \Wbar (-i E_{- \alpha_{IN}})~, \qquad \textrm{where} \cr
\AA =&~  \half (\pm 1 - \cos{\theta}) \ed\phi~, \qquad W = \frac{ (a_{I,N-1} + b) r}{2 \sinh\left( (a_{I,N-1} + b) r \right)} e^{\pm i\phi} (-i \ed\theta + \sin{\theta} \ed\phi)~,
\end{align}
where $a_{I,N-1} = a_I + \cdots + a_{N-1}$ for $I < N$ and is equal to zero if $I = N$.  In this case $\alpha_{IN} \in \Delta_{\rm heavy}$ so the $W$-bosons are eliminated by the projection.  The other new element we need is
\begin{align}\label{projcharge}
\Pi(H_{IN}) =&~ \Pi(H_I) + \cdots + \Pi(H_N) = \sum_{J=I}^N H_J - \frac{N+1}{N} \sum_{J=1}^N C^{NJ} H_J \cr
=&~ \sum_{\check{J} = I}^{N-1} H_{\check{J}}  - \sum_{\check{J} = 1}^{N-1} \check{C}^{(N-1) \check{J}} H_{\check{J}} = H_{I(N-1)} - \check{h}^{N-1}~,
\end{align}
where $H_{I(N-1)}$ is zero if $I = N$.  Terms involving $H_N$ have cancelled out, so $\Pi(H_{IN})$ only depends on the first $N-1$ simple co-roots, which we identify with the simple co-roots of $\mathfrak{su}(N)$: $H_{\check{I}} = \check{H}_{\check{I}}$.  Thus the projected fields are
\begin{align}
 \Pi(X) =&~  \left[ \frac{a_{I,N-1} + b}{2}  \coth\left((a_{I,N-1} + b) r\right) - \frac{1}{2 r} - \frac{a_{I,N-1} + b}{2} \right] \Pi(H_{IN}) + \check{X}_{\infty}~, \cr
\Pi(A) =&~ \AA \ \Pi(H_{IN})~.
\end{align}
The $b\to \infty$ limit exists and is given by
\begin{align}
\lim_{b \to \infty} \Pi(X) =&~ - \frac{1}{2 r} \Pi(H_{IN}) + \check{X}_\infty~, \qquad \lim_{b\to\infty} \Pi(A) = \AA \ \Pi(H_{IN})~.
\end{align}
This is a Cartan-valued field configuration with line defect boundary conditions, with \tHooft and asymptotic magnetic charge
\begin{equation}
\check{P} = \check{\gamma}_{\rm m} = \Pi(H_{IN})~.
\end{equation}

Let us apply the dimension formula \eqref{dim2}:
\begin{equation}\label{dimSUNcase2}
\dim_{\mathbb{R}} T_{[\hat{A}]} \fMM = 2 \sum_{\check{\alpha} \in \check{\Delta}^+} \left( \langle \check{\alpha} , \check{\gamma}_{\rm m} \rangle +| \langle \check{\alpha}, \check{P} \rangle | \right) = \sum_{ \mathclap{\substack{  \check{\alpha} \in \check{\Delta}^+ \\ \langle \check{\alpha}, \check{P} \rangle > 0 }} } 4 \langle \check{\alpha}, \check{P} \rangle ~.
\end{equation}
We compute the pairing $\langle \check{\alpha},\check{P} \rangle$ for a generic positive root $\check{\alpha} = \check{\alpha}_{\check{J}\check{K}}$, $1 \leq \check{J} \leq \check{K} \leq N-1$, using the identity $[i \check{P}, \check{E}_{\check{\alpha}}] = \langle \check{\alpha}, \check{P} \rangle \check{E}_{\check{\alpha}}$ and working in the fundamental representation.  From \eqref{projcharge}, the matrix components of $\check{P}$ in the fundamental representation are obtained from those of the co-root $\check{H}_{I(N-1)}$ and the magnetic weight $\check{h}_{N-1}$.  The co-root $i \check{H}_{I(N-1)}$ is a diagonal matrix with zeros everywhere except for a one on the $I^{\rm th}$ diagonal entry and a minus one on the $N^{\rm th}$ diagonal entry.  The magnetic weight has the form $i\check{h}^{N-1} = N^{-1} \diag(1,\ldots,1,-(N-1))$.  Thus
\begin{equation}
(i \check{H}_{I(N-1)})_{mn} = \delta_{Im} \delta_{In} - \delta_{Nm} \delta_{Nn}~, \qquad (i \check{h}^{N-1})_{mn} =   \frac{1}{N} \delta_{mn} - \delta_{Nm} \delta_{Nn}~.
\end{equation}
The matrix representation of the raising operator $\check{E}_{\check{\alpha}_{\check{J}\check{K}}}$ corresponding to the root $\check{\alpha}_{\check{J}\check{K}}$ has a single non-zero entry in the $\check{J}$-$(\check{K}+1)$ slot: $(\check{E}_{\check{\alpha}_{\check{J}\check{K}}})_{mn} = \delta_{\check{J}m} \delta_{(\check{K}+1)n}$.  A short computation shows
\begin{align}\label{fundtrick}
\left( [ i \check{H}_{I(N-1)}, \check{E}_{\check{\alpha}_{\check{J}\check{K}}} ] \right)_{mn} =&~ (\delta_{I\check{J}} - \delta_{I(\check{K}+1)} + \delta_{(\check{K}+1)N} ) \delta_{\check{J}m} \delta_{(\check{K}+1)n} ~, \cr
\left( [ i \check{h}^{N-1}, \check{E}_{\check{\alpha}_{\check{J}\check{K}}} ] \right)_{mn} =&~ ( \delta_{(\check{K}+1)N} ) \delta_{\check{J}m} \delta_{(\check{K}+1)n} ~.
\end{align}
Hence
\begin{equation}\label{pair1}
\langle \check{\alpha}_{\check{J}\check{K}}, \check{H}_{I(N-1)} \rangle = (\delta_{I\check{J}} - \delta_{I(\check{K}+1)} + \delta_{(\check{K}+1)N} )~, \qquad \langle \check{\alpha}_{\check{J}\check{K}}, \check{h}^{N-1} \rangle =  \delta_{(\check{K}+1)N}~,
\end{equation}
and therefore
\begin{equation}\label{pair2}
\langle \check{\alpha}_{\check{J}\check{K}}, \check{P} \rangle = \delta_{I\check{J}} - \delta_{I(\check{K}+1)} ~.
\end{equation}

We are interested in those $\check{\alpha}_{\check{J}\check{K}}$ for which this is a \emph{positive} quantity.  This will be the case if and only if $\check{J} = I$.  (Note it is not possible to have $\check{J} = \check{K}+1$.)  In particular, if $I = N$ there are no roots for which this is the case, while if $I < N$ there are precisely $N - I$ such roots: $\check{\alpha}_{I \check{K}}$, with $\check{K} \in \{I,\ldots,N-1 \}$.  Each of these roots contributes four to the dimension, \eqref{dimSUNcase2}, so we conclude that
\begin{equation}
\dim_{\mathbb{R}} T_{[\hat{A}]} \fMM = 4 (N - I)~.
\end{equation}
In particular if $I = N$ the dimension is zero.  This result is consistent with the brane picture, Figure \ref{fig6}-$(2b)$, where we see that there are $N-I$ D1-string segments that can move independently.

\subsection{The generic configuration}

\begin{figure}
\begin{center}
\includegraphics{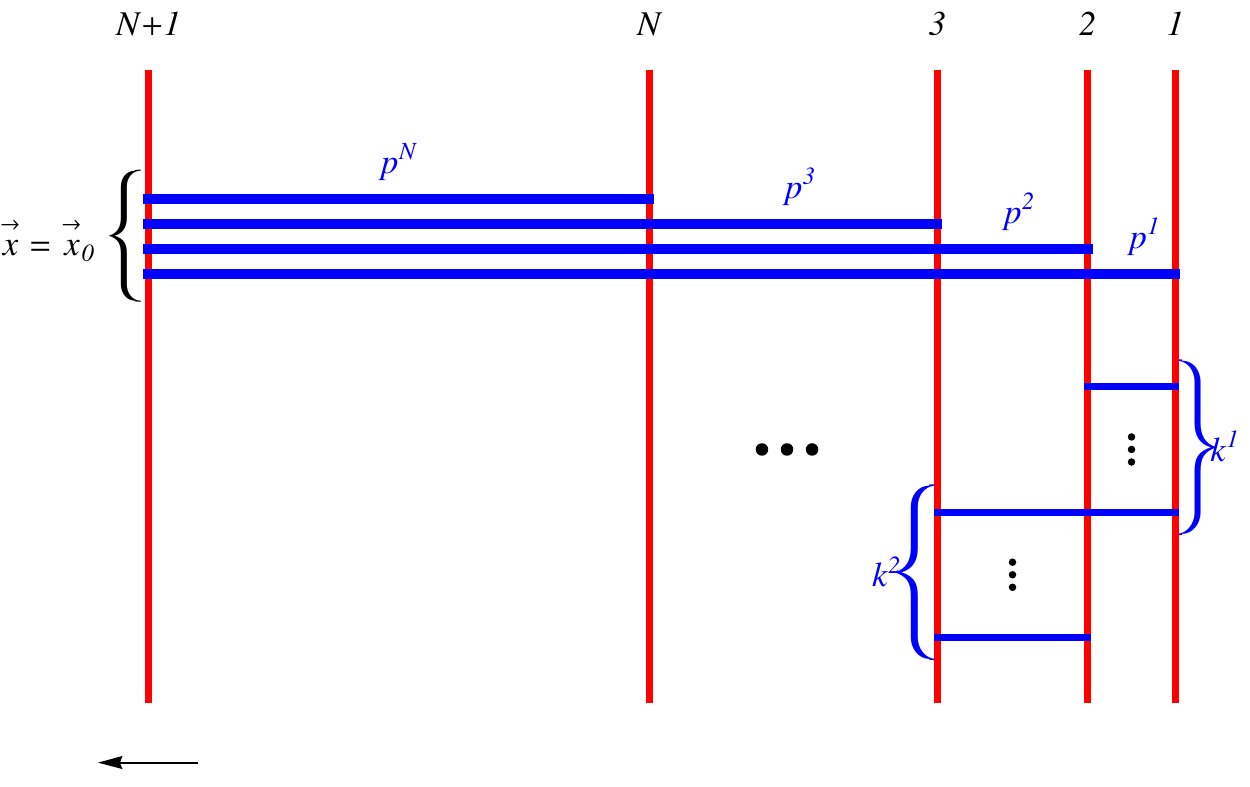}
\caption{The general configuration involving a single \tHooft defect at $\vec{x}_0$ with generic \tHooft charge, and some number of smooth monopoles.  The D1-strings comprising the defect have been artificially separated for clarity, but should be thought of as occupying the same point in $\mathbb{R}^3$.}
\label{fig7}
\end{center}
\end{figure}

Finally let us consider the most generic configuration resulting in a single `t Hooft defect and an arbitrary number of smooth monopoles of each type.  In the $\mathfrak{su}(N+1)$ theory we let $p^1$ D1-strings stretch between the first and last D3-brane, $p^2$ between the second and last, \etc, up to $p^N$ stretched between the last two D3-branes.  All of these strings are taken to be coincident with respect to the $\mathbb{R}^3$ spanned by D3-branes.  When we send the $x^4$-position of the last D3-brane to $-\infty$ they will be responsible for creating the \tHooft defect.  We also let $k^{\check{I}}$ additional D1-strings stretch between the $\check{I}^{\rm th}$ and $(\check{I} + 1)^{\rm th}$ D3-brane at arbitrary positions in $\mathbb{R}^3$, for $1 \leq \check{I} \leq N-1$.  See Figure \ref{fig7}.

Although we cannot write down the explicit fields corresponding to this brane configuration, it is clear that the projection and limiting procedure will produce a singular monopole configuration with one \tHooft defect in the $PSU(N)$ theory.  The \tHooft charge of the defect can be obtained by applying the projection to the magnetic charge generated by the D1-strings that are becoming semi-infinite.  Let us denote this charge $\gamma_{\rm m}^{\rm si}$; it is given by
\begin{equation}
\gamma_{\rm m}^{\rm si} = \sum_{I=1}^N p^I H_{IN}~,
\end{equation}
where recall $H_{IN} = H_I + H_{I+1} + \cdots H_{N}$.  The projection, using \eqref{projcharge}, is
\begin{equation}
\check{P} = \Pi(\gamma_{\rm m}^{\rm si}) = \left\{ \sum_{\check{I} = 1}^{N-1} p^{\check{I}} \left( H_{\check{I} (N-1)} - \check{h}^{N-1} \right) \right\} - p^N \check{h}^{N-1} ~,
\end{equation}
or in other words,
\begin{equation}\label{gentHooftcharge}
\check{P} = \sum_{\check{I} = 1}^{N-1} (p^{\check{I}} - \bar{p} ) \check{H}_{\check{I}(N-1)}~, \qquad \textrm{where} \quad \bar{p} := \frac{1}{N} \sum_{I=1}^N p^I ~.
\end{equation}
The asymptotic charge will be the projection of the total initial charge, which is the semi-infinite charge plus the extra $\sum_{\check{I}} k^{\check{I}} H_{\check{I}}$.  Since $\Pi(H_{\check{I}}) = H_{\check{I}}$, $1 \leq \check{I} \leq N-1$, we have
\begin{equation}\label{gmgenconfig}
\check{\gamma}_{\rm m} = \check{P} + \sum_{\check{I} = 1}^{N-1} k^{\check{I}} H_{\check{I}}~.
\end{equation}

Knowing the charges is sufficient to determine the dimension of the moduli space (since $\check{X}_\infty$ is in the fundamental Weyl chamber by construction).  We have
\begin{equation}
\dim_{\mathbb{R}} T_{[\hat{A}]} \fMM = 2 \sum_{\check{\alpha} \in \check{\Delta}^+} \left( \langle \check{\alpha} , \check{\gamma}_{\rm m} \rangle +| \langle \check{\alpha}, \check{P} \rangle | \right) = 4 \sum_{\check{I}=1}^{N-1} k^{\check{I}} + \sum_{ \mathclap{ \substack{  \check{\alpha} \in \check{\Delta}^+ \\ \langle \check{\alpha}, \check{P} \rangle > 0 }} } 4 \langle \check{\alpha}, \check{P} \rangle ~.
\end{equation}
Now consider $\langle \check{\alpha}_{\check{J}\check{K}}, \check{P} \rangle$ for some root with $1 \leq \check{J} \leq \check{K} \leq N - 1$.  With the aid of \eqref{pair1} and \eqref{pair2} we find
\begin{equation}\label{genrootgenP}
\langle \check{\alpha}_{\check{J}\check{K}}, \check{P} \rangle = p^{\check{J}} - p^{\check{K}+1}~.
\end{equation}
Four times this quantity contributes to the dimension, but only if it is positive.  We have
\begin{equation}\label{gendefectdim}
\dim_{\mathbb{R}} T_{[\hat{A}]} \fMM =  4 \sum_{\check{I}=1}^{N-1} k^{\check{I}} + 2 \sum_{\check{J} = 1}^{N-1} \sum_{\check{K} = \check{J}}^{N-1} \left( p^{\check{J}} - p^{\check{K}+1} + |p^{\check{J}} - p^{\check{K}+1}| \right)~.
\end{equation}

This is our final result for the generic configuration and it is perhaps somewhat unexpected.  In particular, to determine the contribution to the dimension from the D1-strings associated with the defect, we do not simply count the number of finite length segments that could naively be broken off based on the picture\footnote{This would have been $\sum_{\check{I}} (N-\check{I}) p_{\check{I}}$.} of Figure \ref{fig7}.  Rather, for each ordered pair of D3-branes, labeled by $(\check{J}, \check{K}+1)$, we must compare the number of D1-strings originating on the $\check{J}^{\rm th}$ brane and the number originating on the $(\check{K} +1)^{\rm th}$ brane.  If the former is larger then the difference adds to the quaternionic dimension of the moduli space, while if they are equal or the latter is larger, there is no contribution to the dimension.

Although this seems at first puzzling from the point of view of the brane picture, we note that the result is perfectly consistent with our analysis of the Cartan-valued solutions in \cite{MRVdimP1}.  As an extreme example, consider the case where $p^1 \leq p^2 \leq \cdots \leq p^N$.  Then we would get no contribution to the dimension from the $p^I$ since $\langle \check{\alpha}, \check{P} \rangle \leq 0$ for all positive roots $\check{\alpha}$.  However this is precisely what we expect from \cite{MRVdimP1} since such an \tHooft charge is in the anti-fundamental Weyl chamber, and corresponds to a ``pure'' defect.  In the following section we interpret this result from the point of view of the branes.  We are naturally led to the notions of monopole extraction and bubbling.

\section{Stuck branes, monopole extraction, and monopole bubbling}\label{sec:stuck}

\begin{figure}
\begin{center}
\includegraphics{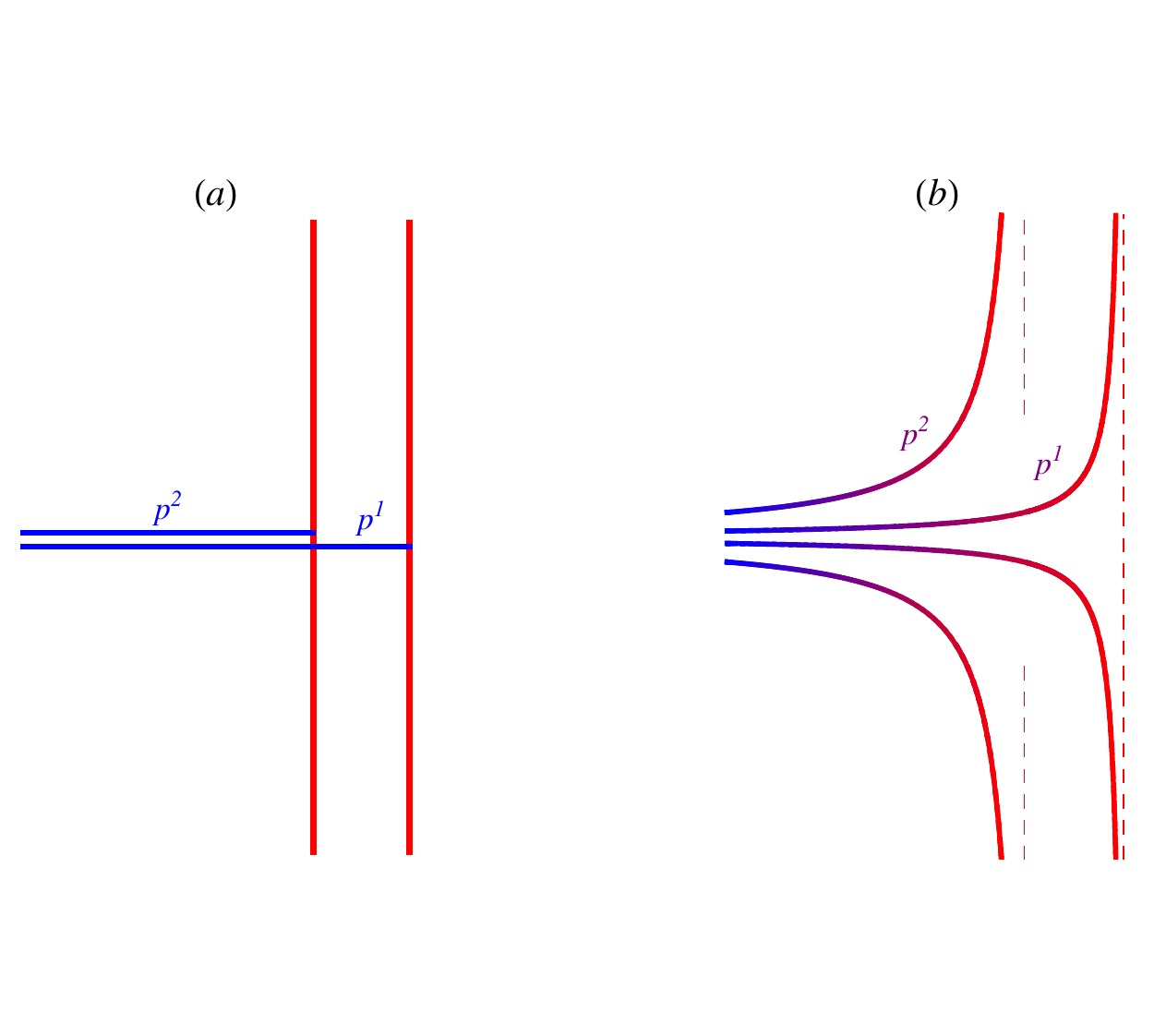}
\caption{$(a)$ D1-string idealization of a $\mathfrak{u}(2)$ \tHooft defect with charges $(p^1,p^2) = (1,3)$.  $(b)$ The actual bending of the D3-branes as determined by the Higgs field configuration.}
\label{fig8}
\end{center}
\end{figure}

In this section we are only considering the reduced theory, after the projection and limiting procedure, so we will drop the convention of putting $\check{}$ 's on everything.  Let us consider the simplest scenario in which the naive counting of mobile brane segments gives a different answer from the dimension formula: two D3-branes with $p^1$ D1-strings coming in from $x^4 = -\infty$ and ending on the right brane and $p^2 \geq p^1$ D1-strings ending on the left brane.  See Figure \ref{fig8}-$(a)$.  Our methods above indicate that this configuration has asymptotic magnetic charge and \tHooft charge
\begin{equation}
\gm = P = p^1 H - (p^1 + p^2) h = (p^1- p^2) h~,
\end{equation}
where we used the relation $H = 2 h$ between the co-root and magnetic weight of the $PSU(2)$ theory.  For any $p^2 \geq p^1$ this is in the closure of the anti-fundamental Weyl chamber and hence the dimension formula gives zero for the dimension of the moduli space.

We are led to the interpretation that the $p^1$ D1-string segments stretching between the two branes are ``stuck,'' and cannot break off from their semi-infinite continuations on the other side of the left D3-brane.\footnote{At the end of this section we will comment on the possibility of ``un-sticking'' stuck branes through the process of monopole bubbling.}  Intuition for this phenomenon is obtained by recalling that these pictures of orthogonal rigid branes are an idealization of what really happens.  In this simple situation we can understand the brane bending precisely by writing down the Cartan-valued Higgs field.  For the purposes of visualization it is better to work with the $\mathfrak{u}(2)$-valued configuration, which takes the form
\begin{equation}
X = \left( \begin{array}{c c} x_1 - \frac{p^1}{2r} & 0 \\ 0 & x_2 - \frac{p^2}{2r} \end{array} \right)~.
\end{equation}
This corresponds to the brane bending shown in Figure \ref{fig8}-$(b)$; $x_1 > x_2$ as in the figure.  Observe that the world-volume of the right brane never touches that of the left brane as long as $p^1 \leq p^2$.  Unlike the picture shown in Figure \ref{fig1}, the two branes never meet.  The $p^1$ semi-infinite D1-strings originating on the right D3-brane never intersect the left D3, but rather extend down the throat created by the $p^2$ strings originating on the left brane.\footnote{The more accurate throat---or ``BIon''---picture \cite{Gibbons:1997xz,Callan:1997kz} of Figure \ref{fig8}-$(b)$ can be reproduced from the D1-string point of view by considering the non-Abelian DBI action for multiple D1-strings \cite{Constable:1999ac}.}  A similar picture was employed by Gaiotto and Witten in Figure 11 of \cite{Gaiotto:2008sa}.

\begin{figure}
\begin{center}
\includegraphics{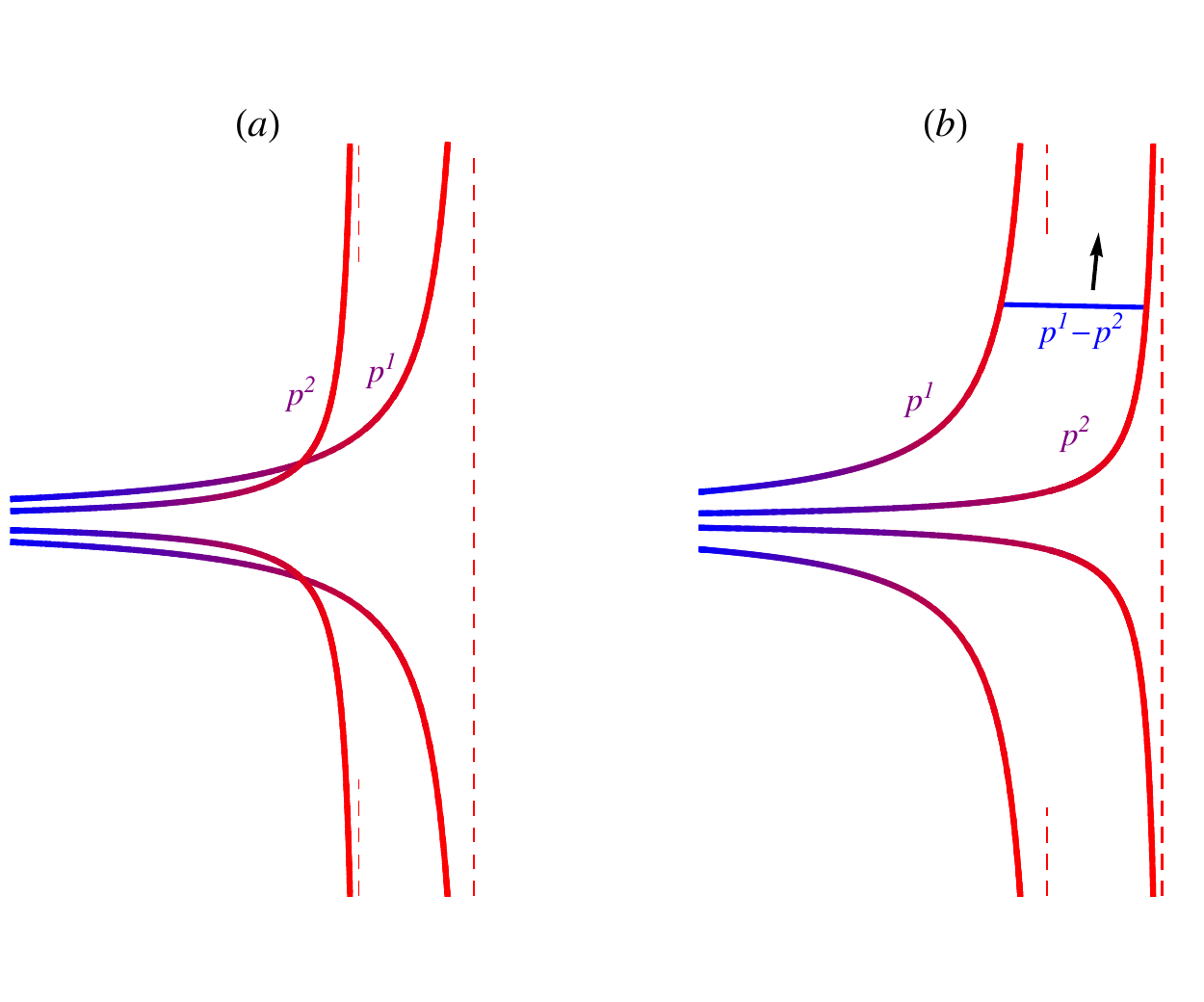}
\caption{$(a)$ When $p^1 > p^2$, the brane worldvolumes intersect.  $(b)$ Infinitesimal length D1-strings can be created at the intersection and a maximum number of $p^1 - p^2$ can escape the throat region.}
\label{fig9}
\end{center}
\end{figure}

Now consider the situation $p^1 > p^2$, so that the dimension formula gives a dimension for the moduli space of $4(p^1-p^2)$.  When $p^1 > p^2$ the brane world-volumes collide; see Figure \ref{fig9}-$(a)$.  In this region the local value of the $\mathfrak{su}(2)$ Higgs field (the separation between the branes) is small and massless monopoles can be produced, represented by infinitesimal length D1-strings connecting the D3-branes.  The dimension formula suggests that $p^1 - p^2$ D1-strings should be able to escape the throat region and become macroscopic as depicted in Figure \ref{fig9}-$(b)$.  The resulting configuration would have $p^1 - p^2$ finite length D1 segments which could move freely and independently and an \tHooft defect with $U(2)$ charge $({p'}^1,{p'}^2) = (p^2,p^1)$, or $PSU(2)$ charge $P' = \half (p^2 - p^1) H$.  Since $p^2 < p^1$ this new \tHooft charge is in the anti-fundamental Weyl chamber and thus there are no degrees of freedom associated with the defect; all are accounted for by the smooth mobile monopoles.

Figure \ref{fig9}-$(b)$ is of course only a cartoon of what really happens; when the $p^1 - p^2$ D1-strings are in the throat region it would be more appropriate to represent them as some localized bending of branes.  However there are good reasons to think that this picture is qualitatively correct.  The initial and final \tHooft charges of the defect are related by a Weyl reflection.  Thus, by making a local gauge transformation on the field configuration representing Figure \ref{fig9}-$(a)$, we can view it as a field configuration with the same defect charge as Figure \ref{fig9}-$(b)$.  The asymptotic magnetic charge stays the same throughout the process, and hence the regularized energy stays the same.  The energy required to produce the massive monopoles as they leave the throat region is obtained by adjusting the shape of the throat.  Since the \tHooft and asymptotic data are equivalent, these configurations represent two points on the \emph{same} moduli space $\fMM(P;\gm; X_\infty)$.  If the moduli space is connected as we assume, then there must be a path with these two configurations as its endpoints.\footnote{Again, since the branes depicted in Figure \ref{fig9}-$(b)$ are merely a cartoon of a Higgs field configuration solving the Bogomolny equation, we can not say precisely which point in moduli space this figure corresponds to.  It is thus perhaps more accurate to say that this figure is representative of some region in the moduli space.  This is in contrast to to Figure \ref{fig9}-$(a)$ which represents the Higgs field profile of an actual solution, namely a Cartan-valued solution, and hence represents a specific point in moduli space.}

\begin{figure}[t]
\begin{center}
\includegraphics{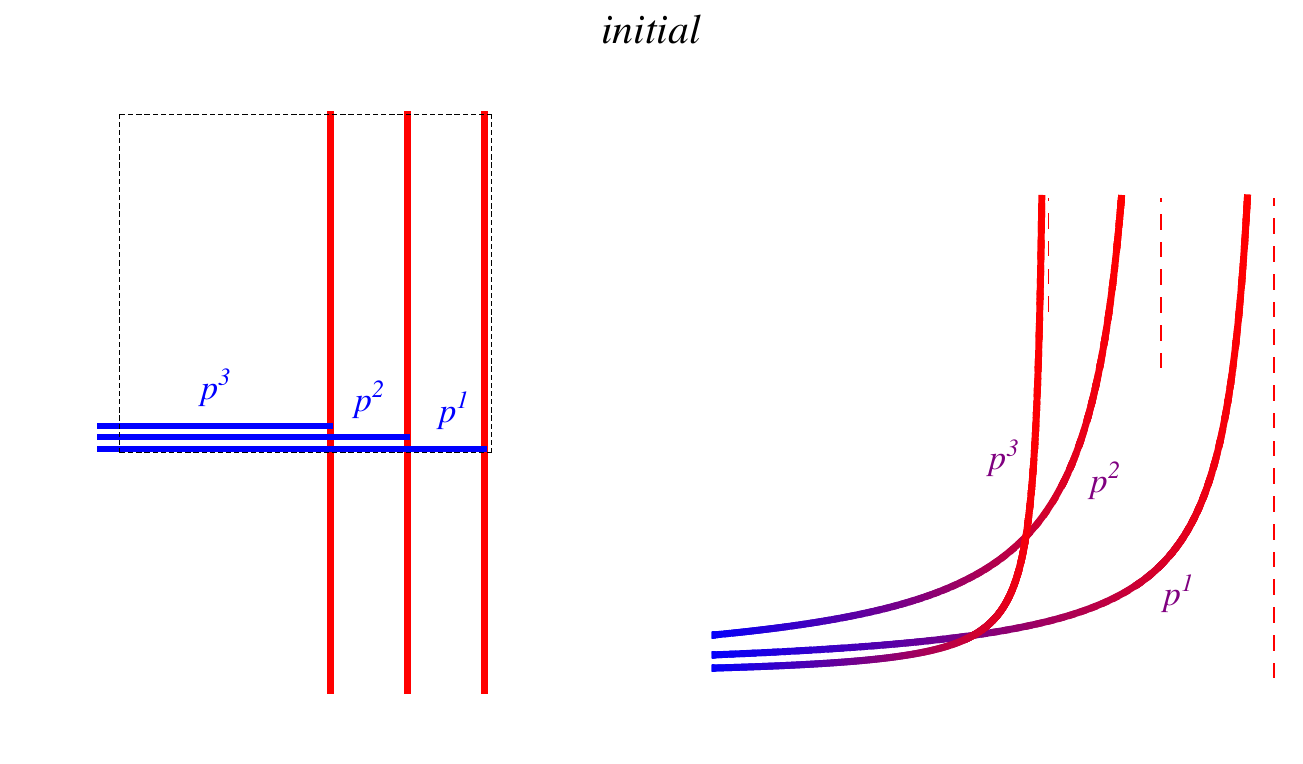}
\includegraphics{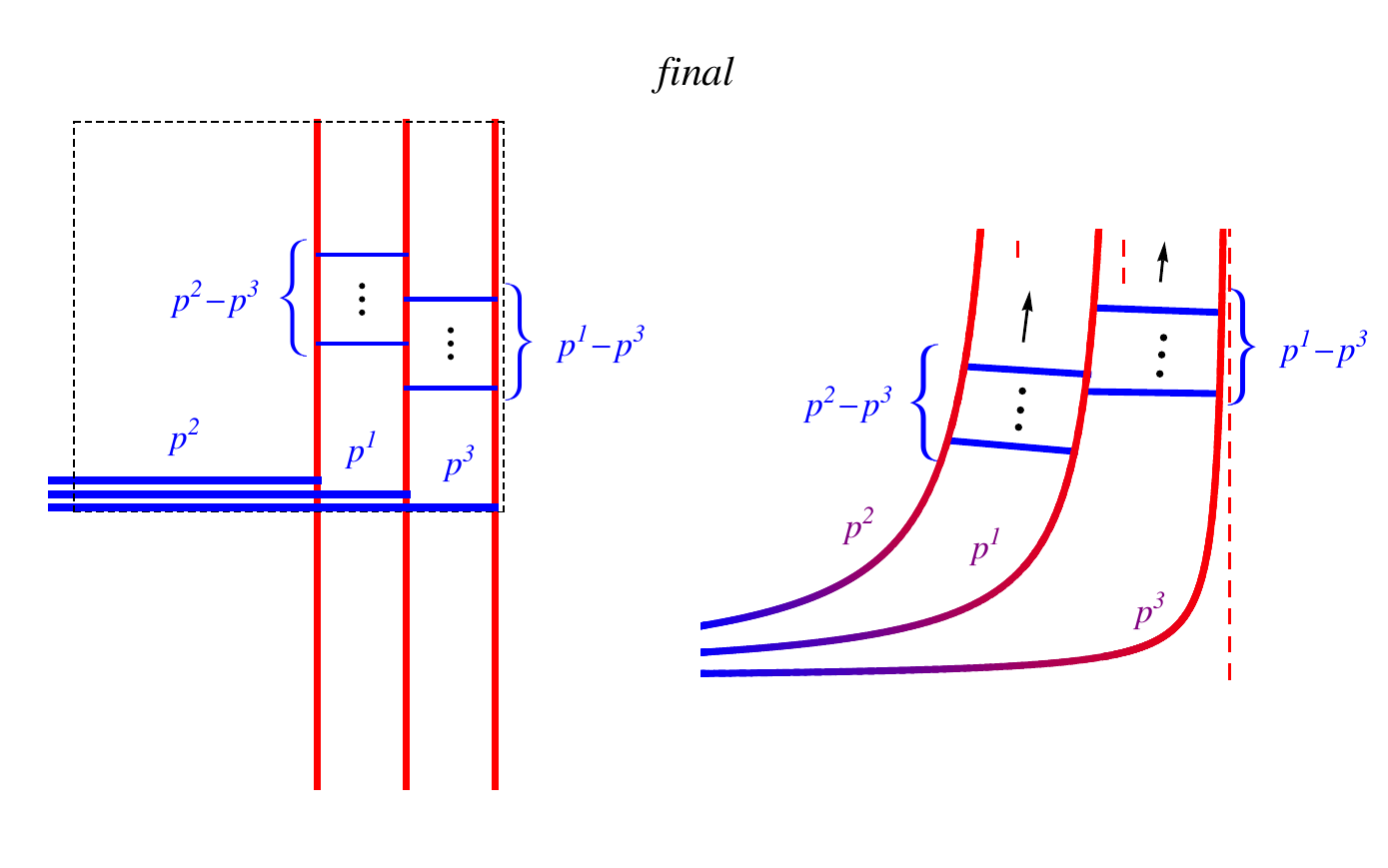}
\caption{A $\mathfrak{u}(3)$ example with defect charges $(p^1,p^2,p^3) = (4,6,1)$.  Two different sequences of brane motion that lead from the initial configuration to the final one can be found in Figure \ref{fig11} below.}
\label{fig10}
\end{center}
\end{figure}

\begin{figure}
\begin{center}
\includegraphics{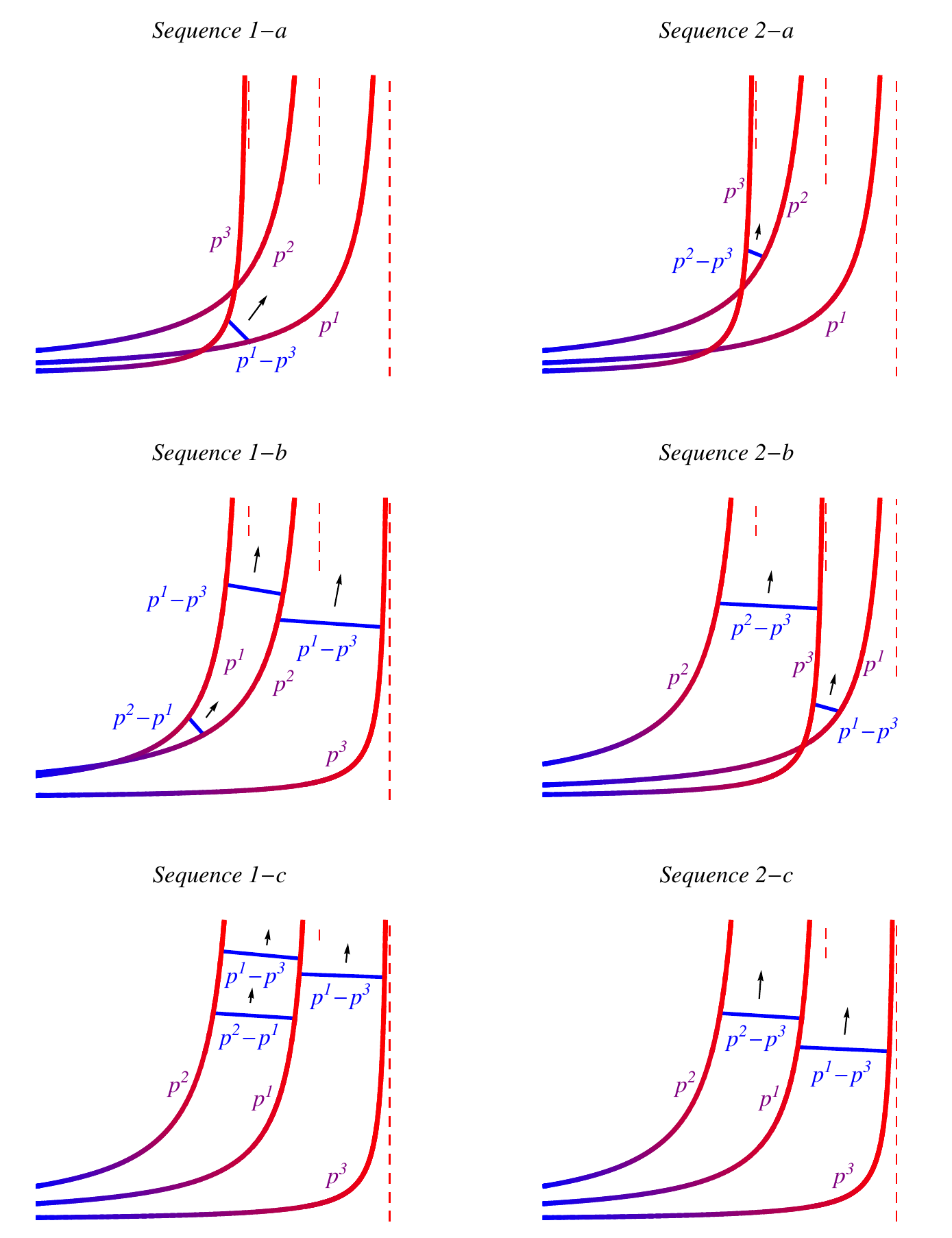}
\caption{Two sequences by which the maximal number of monopoles can be extracted from the $\mathfrak{u}(3)$ defect of Figure \ref{fig10}.  Either way, in the final configuration we end up with a pure \tHooft defect, $p^1-p^3$ D1-strings stretched between the first and second D3-brane, and $p^2-p^3$ D1-strings stretched between the second and third D3-brane.  Note in sequence $1$-$(a)$ the $p^1-p^3$ D1-strings should be thought of as infinitesimal strings located at the intersection locus of the left-most and right-most brane.  We have displaced them slightly from this locus for clarity.  Similar remarks apply for sequences $1$-$(b)$, $2$-$(a)$, and $2$-$(b)$.}
\label{fig11}
\end{center}
\end{figure}

In Figure \ref{fig10} we consider a more complicated example with an initial $\mathfrak{u}(3)$ defect of charges $(p^1,p^2,p^3)$ with $p^2 > p^1 > p^3$.  According to \eqref{gendefectdim}, the quaternionic dimension of the moduli space receives contributions from two pairs and is given by $(p^1-p^3) + (p^2 - p^3)$.  In Figure \ref{fig11} we depict two different processes by which this number of finite length D1 segments can be emitted from the defect.  The resulting defect charge is described by $({p'}^1,{p'}^2,{p'}^3) = (p^3,p^1,p^2)$ and corresponds to the Weyl representative of the initial \tHooft charge in the anti-fundamental Weyl chamber.  Again this process can be described by motion on a moduli space $\fMM( [P]; \gm; X_\infty)$.\footnote{Since we defined the moduli space in \eqref{Mdef} to depend on $P$ and not the Weyl orbit, we should make a local gauge transformation at each stage to a fixed Weyl representative, say $P^-$, in order to describe this process as motion on a fixed moduli space.}  Depending on which path we follow we pass through different Weyl representatives of $[P]$ at intermediate stages.

This example can clearly be generalized to understand the general form of \eqref{gendefectdim}.  Consider an \tHooft defect $P$ with corresponding $\mathfrak{u}(N)$ charges $(p^1,\ldots p^N)$ as in Figure \ref{fig7}.  Any ordered pair of branes with charges $(p^I,p^J)$ such that $I < J$ and $p^I > p^J$ will have intersecting worldvolumes.  At such an intersection $p^I - p^J$ finite length D1-strings can be created and moved away from the throat region.  We call this \emph{monopole extraction}.  After all available D1-strings have moved away the remaining defect will be pure with charge in the anti-fundamental Weyl chamber, $P' = P^-$.  In terms of $U(N)$ charges, $({p'}^1,\ldots,{p'}^N)$ will be an increasing sequence of integers.  This \tHooft charge does not contribute to the dimension, and in the final configuration \eqref{gendefectdim} is fully accounted for by the mobile branes.  The configurations considered in case 3 of Figure \ref{fig5} and case 2 of Figure \ref{fig6} can be viewed as special cases where the initial charges are $(p^1,p^2) = (1,0)$ and $p^{J} = {\delta_I}^J$ respectively.

Now we can also understand the motivation for the conjecture stating when $\fMM$ is non-empty, at least in the case of $\mathfrak{g} = \mathfrak{su}(N)$.  Recall that this is when the relative charge, $\tilde{\gamma}_{\rm m} = \gm - \sum_n P_{n}^-$, is a sum of simple co-roots with non-negative coefficients: $\tilde{\gamma}_{\rm m} = \sum_I \tilde{m}^I H_I$, with $\tilde{m}^I \geq 0, \forall I$.  After taking into account the effects of brane bending in the vicinity of the defects as described above, we see that the coefficients $\tilde{m}^I$ are precisely the number of mobile branes between the $I^{\rm th}$ and $(I+1)^{\rm th}$ D3-brane.

The processes we have described \emph{thus far} should not be viewed as monopole bubbling.  In the phenomenon of monopole bubbling an \tHooft defect emits or absorbs smooth monopoles, changing its charge in a physical way---\ie\ not by a Weyl transformation.  The moduli spaces before and after the bubbling are different; in particular their dimension is different.    The picture of stuck branes we have developed can also be considered in such situations, and this leads to some qualitative understanding of monopole bubbling, as we now explain.  A similar description of monopole bubbling within the context of the D3-D1 system has been given in \cite{Gang:2012yr}.

\clearpage

\begin{figure}
\begin{center}
\includegraphics{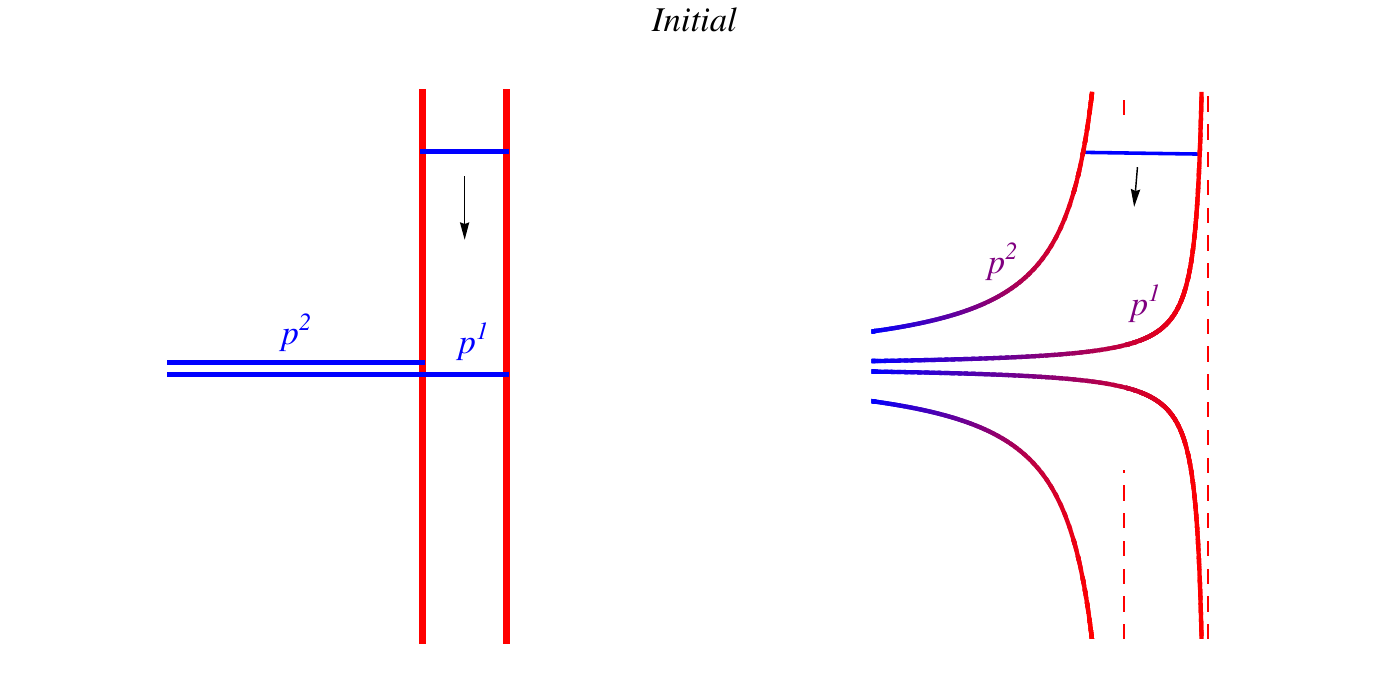}
\includegraphics{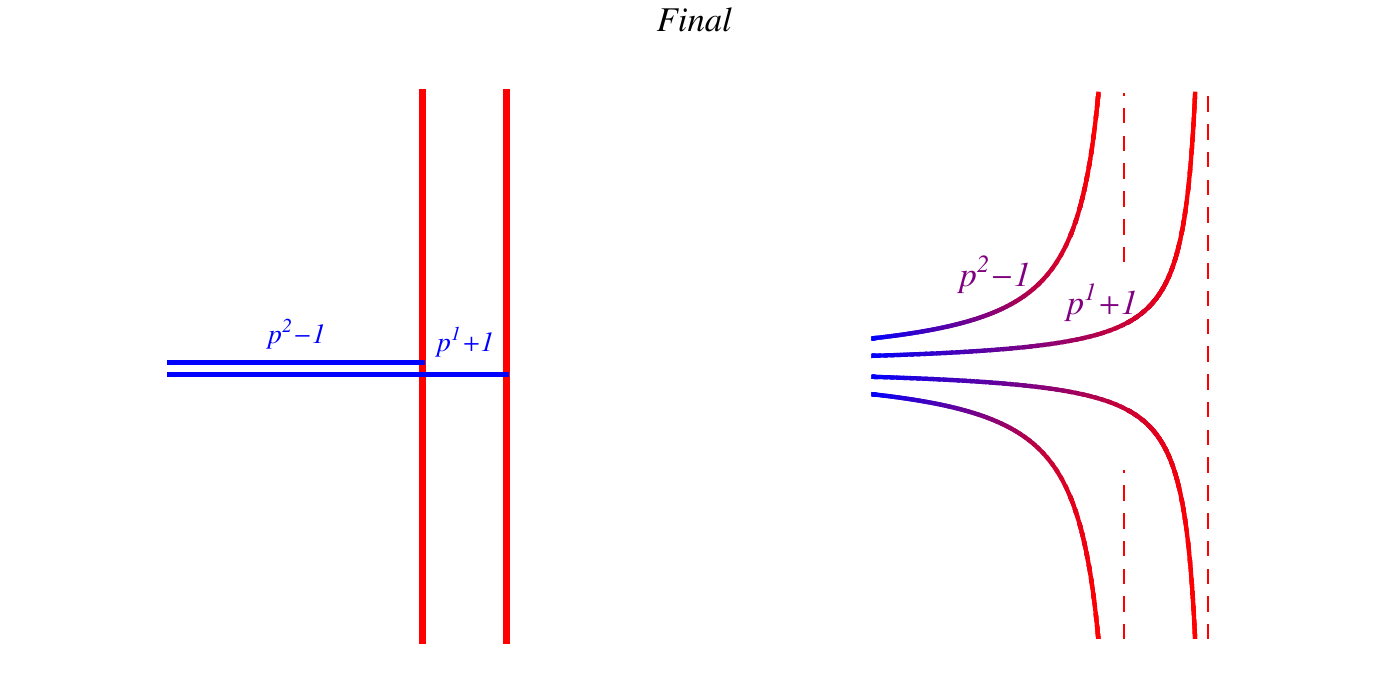}
\caption{An example of monopole bubbling in which an \tHooft defect of charge $(p^1,p^2)$ absorbs a smooth $\mathfrak{su}(2)$ monopole, resulting in a new \tHooft defect of charge $({p'}^1,{p'}^2) = (p^1 + 1, p^2 -1)$.  In the process one moveable brane joins to one of the $p^2$ semi-infinite branes ending at $x_2$ to become a semi-infinite brane ending at $x_1$.}
\label{fig12}
\end{center}
\end{figure}

Let us return to the $U(2)$ system of Figure \ref{fig8} and let us assume that $p^2 \geq p^1 + 2$.  Suppose there is a finite length D1-string representing a smooth monopole some distance from the defect and we send it in towards the defect.  With the D1-string at any finite separation from the defect we expect a four-dimensional moduli space.  However if the D1-string is moved directly on top of the defect, we get a new configuration consisting of a pure \tHooft defect with charge $({p'}^1,{p'}^2) = (p^1 - 1, p^2+1)$.  The process is depicted in Figure \ref{fig12}.  This corresponds to a new $\mathfrak{su}(2)$ charge that is still in the closure of the anti-fundamental Weyl chamber and hence the dimension of the new moduli space is zero.  The D1-string we sent in has become stuck.  The defect charge has changed and as a consequence the dimension has changed in a way that is consistent with the description of monopole bubbling given in \cite{Kapustin:2006pk}.

This picture indicates that the lower-dimensional moduli space, in this case a point, fills in a locus that has been cut out of the higher-dimensional moduli space.  The motion of the D1-string towards the defect can be described as motion towards a boundary of the four-dimensional space created by the removal of a zero-dimensional submanifold.  It was found in \cite{Kapustin:2006pk} that the lower-dimensional space could be described as the fixed point locus of an orbifold action on the higher-dimensional space.  In the simplest case of one smooth $\mathfrak{su}(2)$ monopole in the presence of an $SO(3)$ defect with twice the minimal charge, $(p^1,p^2) = (0,2)$, (or more generally $(p^1,p^2) = (p^1,p^1+2)$) one can verify this picture explicitly using results of Cherkis and Kapustin on the moduli space \cite{Cherkis:1997aa,Cherkis:1998hi}.  They found that the moduli space for this example is a degenerate two-centered Taub-NUT manifold, with coincident centers.  Sending the smooth monopole towards the defect corresponds to approaching the nut point, the vicinity of which is an $A_1$ singularity.

In contrast, the moduli space of one smooth $\mathfrak{su}(2)$ monopole in the presence of a minimal $SO(3)$ \tHooft defect, $(p^1,p^2) = (0,1)$, is single-centered Taub-NUT, which is perfectly smooth at the nut.  This is consistent with the brane picture in that the finite-length D1-string does not get stuck in this case: after sending it in we find an \tHooft defect with $({p'}^1,{p'}^2) = (1,0)$ as pictured in Figure \ref{fig5}-$(3b)$.  The field configuration of this defect has a four-dimensional space of zero-modes corresponding to moving the D1-string back off.  The explicit field configurations corresponding to these situations are constructed and compared in \cite{Cherkis:2007jm}.

The general brane picture of Figure \ref{fig7} suggests that in higher dimensional moduli spaces this picture persists and is enhanced: there are different types of orbifold loci corresponding to different types of fundamental monopoles being absorbed by the defect, and there are loci within loci as multiple monopoles are absorbed.  It would be interesting to explore this rich global structure in more detail, but it is beyond the scope of this paper.

Notice in the example of Figure \ref{fig12} that, although the initial and final configurations correspond to points in different singular monopole moduli spaces, the asymptotic charges of the configurations are the same and hence the regularized energy is the same.  This suggests that the process of the smooth monopole being absorbed by the defect, \ie\ the D1-string becoming ``stuck,'' should be reversible.  An \tHooft defect can equally well emit a smooth monopole, changing in the process the Weyl orbit of its charge.  In other words, stuck branes can become un-stuck.  We can understand this possibility from the brane bending point of view as follows.  Although the D3-brane worldvolumes in the right-hand diagrams of Figure \ref{fig12} never collide, they do become arbitrarily close to each other as $x^4 \to -\infty$.  The cost in energy to create a D1-string becomes infinitesimally small as we go all of the way down the throat.  Hence we can think of monopole bubbling as a mobile D1-string coming up the throat from $x^4 = -\infty$.

\section{Wall-crossing of the index and brane motion}\label{sec:branejump}

As we noted below equation \eqref{eq:IndexJump}, the index jumps as a function of $X_\infty$.
In the case of gauge algebra $\mathfrak{g}= \mathfrak{su}(N)$, the brane pictures we have been discussing  can be used to interpret the jumping behavior.  In this case changing $X_\infty$ is the same as changing the asymptotic locations $x_I$ of the D3-branes.

\subsection{Wall-crossing of the index for smooth monopoles}

Let us first consider the case of the adjoint representation, where the index determines the dimension of the moduli space and corresponds to the number of mobile D1-branes.\footnote{Jumping behavior of a related index was observed and analyzed in \cite{Poppitz:2008hr}.}  In this case the index \eqref{indLrho} is
\begin{align}\label{adindex}
\ind{L} =&~ \sum_{\alpha \in \Delta^+} \left\{ \sgn(\langle \alpha, X_\infty \rangle) \langle \alpha, \gm \rangle + \sum_{n=1}^{N_t} | \langle \alpha, P_n \rangle | \right\}~,
\end{align}
twice which gives the dimension of the moduli space $\fMM$, when it is non-empty.  Let us further suppose for the moment that there are no \tHooft charges present, so that $P_n = 0$.  Then the condition for the moduli space to be non-empty is that the components of $\gm$ along the basis of simple co-roots determined by $X_\infty$ are non-negative, and at least one is strictly positive.

The index jumps when $\langle \alpha , X_\infty \rangle$ passes through zero and switches sign for some root $\alpha$.  If we start with a configuration where $X_\infty$ is in the fundamental Weyl chamber---as we have always assumed, and can be arranged by a global gauge transformation---and vary $X_\infty$ continuously, then $\langle \alpha, X_\infty \rangle$ will first go to zero for some simple root $\alpha = \alpha_I$.  The difference in the index after the wall minus before the wall is
\begin{equation}\label{DeltaIndexad}
\Delta_I \ind L = -2 \langle \alpha_I, \gm \rangle = -2 \sum_J C_{IJ} m^J~,
\end{equation}
where we plugged in $\gm = \sum_J m^J H_J$.  We assume the initial magnetic charge $\gm$ is such that the initial moduli space is non-empty.  Then twice \eqref{DeltaIndexad} is the change in dimension provided that the new moduli space is also non-empty.  

To determine whether or not the new moduli space is empty we need to determine the magnetic charge with respect to the new system of positive co-roots obtained from the new $X_\infty$.  The relationship between the new and old $X_\infty$ can conveniently taken to be given by a Weyl reflection about the root $\alpha_I$.  If we set
\begin{equation}
X_{\infty}^{\rm new} = w_{\alpha_I}(X_{\infty}^{\rm old}) := X_{\infty}^{\rm old} - \langle \alpha_I, X_{\infty}^{\rm old} \rangle H_I~,
\end{equation}
this has the required property that $\sgn(\langle \alpha_I, X_{\infty}^{\rm new} \rangle) = - \sgn( \langle \alpha_I, X_{\infty}^{\rm old} \rangle)$ while the signs for all other positive roots remain the same.  This follows from the fact that $C_{IJ} \leq 0$ for $I \neq J$.  Now, if the old and new $X_\infty$ are related by a Weyl reflection, then so are the old and new bases of positive co-roots.  It follow that we can determine the components of the magnetic charge with respect to the new basis by acting with the same Weyl reflection on the magnetic charge.  Here we are simply changing from a passive point of view, where it is the basis of simple co-roots that is changing, to an active point of view where it is the asymptotic magnetic charge that is changing relative to a fixed basis.  (We should therefore use the inverse Weyl transformation on the magnetic charge, but reflections square to one so it is the same.)  Thus we compute
\begin{align}
w_{\alpha_I}(\gm) = \sum_J \left( m^J H_J - C_{IJ} m^J H_I \right) =&~ \left( m^I - \sum_J C_{IJ} m^J \right) H_I + \sum_{J \neq I} m^J H_J  \cr
\equiv&~ \sum_{J} m_{\rm new}^{J} H_J~.
\end{align}
The coefficient of each simple co-root is required to be non-negative, thus we require $m^I \geq \sum_J C_{IJ} m^J$ and at least one $m_J > 0$ for $J \neq I$.  Note if $m_J  = 0$ for all $J \neq I$ then $m^I - \sum_J C_{IJ} m^J = -m^I < 0$, and the new moduli space is empty.  In particular for $\mathfrak{g} = \mathfrak{su}(2)$ gauge theory the new moduli space will always be empty.  We can determine the dimension of the new moduli space directly from the formula $\dim \MM^{\rm new} = \sum_J 4 m_{\rm new}^{J}$ and we find a result consistent with \eqref{DeltaIndexad}:
\begin{align}\label{smoothDdim}
& \dim \MM^{\rm new} = \dim \MM^{\rm old} - 4 \sum_J C_{IJ} m^J~,  \quad  \textrm{if} ~ m^I \geq \sum_J C_{IJ} m^J \quad \& \quad \sum_{J\neq I} m^J > 0~, \cr
&  \MM^{\rm new} = \varnothing~, \qquad \textrm{otherwise}~. \raisetag{24pt}
\end{align}

Specializing to $\mathfrak{su}(N)$ gauge theory, this result has a nice interpretation in terms of brane motion.  Starting with $X_\infty$ in the fundamental Weyl chamber means that we have chosen an ordering of our $N$ D3-branes such that their asymptotic $x^4$-positions satisfy $x_1 > x_2 > \cdots > x_N$.  The distance between the $I^{\rm th}$ and $(I+1)^{\rm th}$ D3-brane is given by $x_{I} - x_{I+1} \equiv a_I = \langle \alpha_I, X_\infty \rangle$.  Thus sending $\langle \alpha_I ,X_\infty \rangle$ through zero corresponds to the $I^{\rm th}$ and $(I + 1)^{\rm th}$ D3-branes passing through each other and exchanging order. 

Monopoles are represented by D1-strings stretching between D3-branes, and these D1-strings carry an orientation that we have so far suppressed.  As we explained above formula \eqref{su2embed}, this orientation is directly related to the sign choice in the Bogomolny equation.  In order for a configuration of D3-branes and D1-strings to represent a solution to the Bogomolny equation, all D1-strings must carry the same orientation.  However, when we pass the $I^{\rm th}$ and $(I + 1)^{\rm th}$ D3-brane through each other, any D1-strings that were initially stretched between them will reverse orientation.  The new configuration with anti-D1's will not represent a solution to the Bogomolny equation.  

\begin{figure}
\begin{center}
\includegraphics{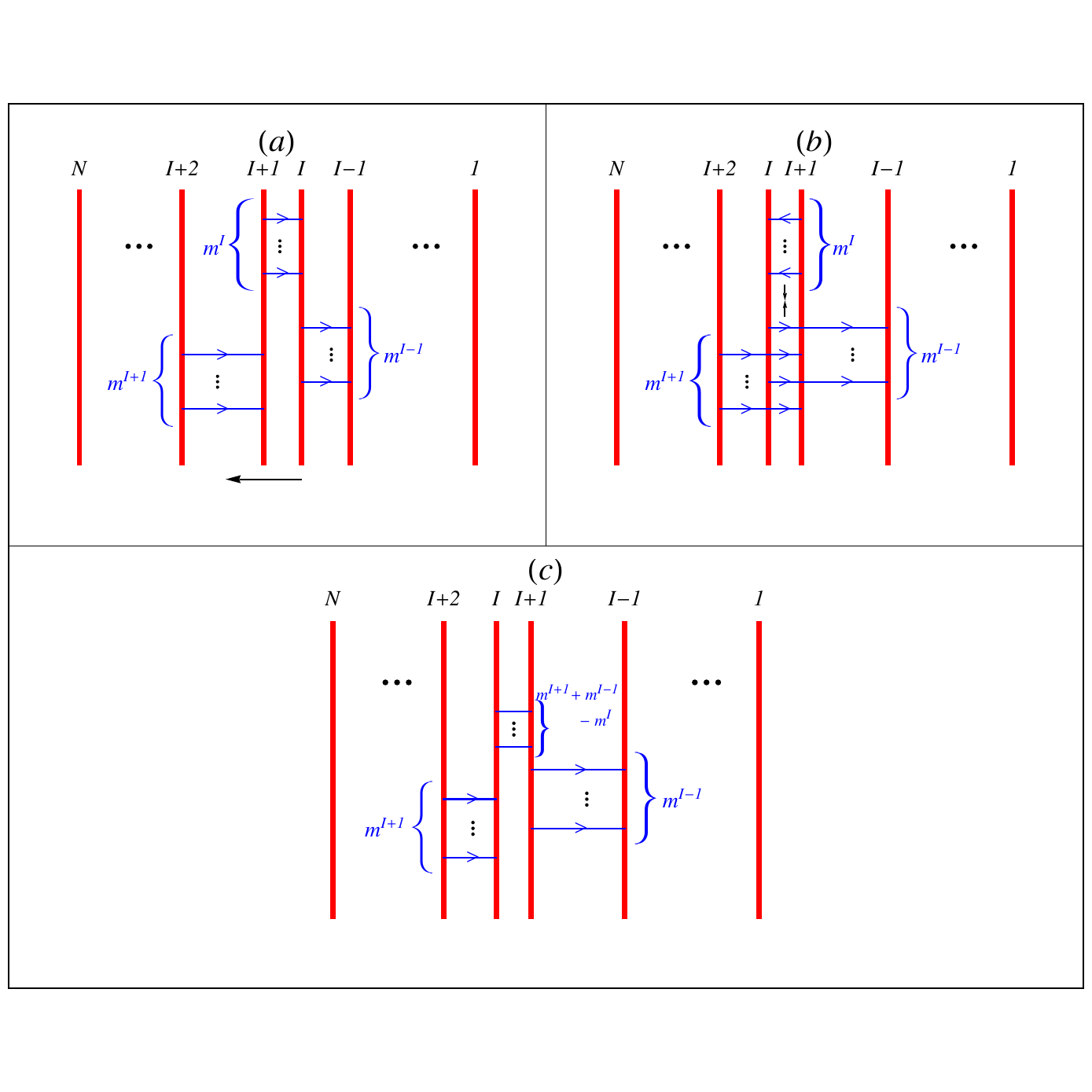}
\caption{Wall-crossing for smooth monopoles in $\mathfrak{g} = \mathfrak{su}(N)$ theory. $(a)$: we start with a generic monopole configuration as in Figure \ref{fig3}, and we move the $I^{\rm th}$ D3-brane to the left, past the $(I+1)^{\rm th}$ D3-brane, as indicated by the arrow. $(b)$: after doing so we find a configuration with both D1-strings and anti-D1's stretched between the $I^{\rm th}$ and $(I + 1)^{\rm th}$ D3-brane; they will attract as indicated by the arrows and annihilate pairwise.  $(c)$: at the end of the process we are left with $m_{\rm new}^{I} = m^{I+1} + m^{I-1} - m^{I}$ D1-strings stretched between this pair of D3-branes.}
\label{fig13}
\end{center}
\end{figure}

What happens next depends on what other types of D1-strings are present.  The expectation from D-brane dynamics is that these new anti-D1's will attract and annihilate D1's of the same type---\ie\ stretched between the same D3-branes so that they carry charge along the same simple co-root.  After the annihilation, if we are left with a system that contains only D1-strings, this will represent a point in the new moduli space, the dimension of which is determined by the number of remaining D1-strings.  If on the other hand we are left with a system that contains any anti-D1's or no D1-strings of either orientation, the new moduli space is empty.  We depict the generic situation in Figure \ref{fig13}.  Using the explicit form of the Cartan matrix for $\mathfrak{su}(N)$, the condition $m_I \geq \sum_J C_{IJ} m^J$ is equivalent to $m_{\rm new}^I = m^{I-1} + m^{I+1} - m^I \geq 0$, and we see that this is the number of remaining D1-strings stretched between the $I^{\rm th}$ and $(I+1)^{\rm th}$ D3-brane at the end of the process.  If the number is negative we interpret its absolute value as the number of anti-D1's.  In that case the new brane configuration does not represent a solution to the Bogomolny equation and the moduli space with the corresonding asymptotic data is empty.

\subsection{Wall-crossing of the index for singular monopoles}

Now we add \tHooft defects to the above story.  Naively, they are represented by semi-infinite D1-strings, but as we demonstrated above, this picture is too crude to account for the dimensions of singular monopole moduli spaces in terms of mobile D1-strings.  We have found that we can understand the dimension in this way once we take into account the effects of brane bending, and replace the semi-infinite D1-strings by BIon spikes.  

Thus it is essential to take into account brane bending---at least the bending corresponding to the defect---when considering wall-crossing phenomena for singular monopoles.  Ideally, one would like to additionally take into account the localized bending due to the finite-length D1-strings, as we described in Figure \ref{fig1}.  However in all but the simplest scenarios this cannot be done precisely since it is tantamount to having explicit solutions of the Bogomolny equation.  One expects that for the case of smooth monopoles corresponding to localized bending, the physics is faithfully represented by the finite-length D1-string idealization.

For singular monopoles it is very instructive to begin by considering the extreme example of the Cartan-valued solutions, where the asymptotic magnetic charge is the \tHooft charge, $\gm = P$.  We will consider a single \tHooft defect; the generalization of the following to multiple defects is straightforward since the Cartan-valued solutions obey a superposition principle.  We consider the process of moving $X_\infty$ across a wall of the fundamental Weyl chamber, where $\langle \alpha_I, X_\infty \rangle$ goes through zero for some simple root $\alpha_I$.  First we determine the change in the index, and hence the change in the dimension of the moduli space, by making use of \eqref{adindex}.  Then we compare with the result obtained from considering the corresponding brane motion.

Let us suppose that the initial configuration has \tHooft charge in the closure of the anti-fundamental Weyl chamber, $\gm = P = P^-$.  This corresponds to an isolated solution of the Bogomolny equation, representing the point of a zero-dimensional moduli space.  Now we send $X_\infty$ to a wall such that $\langle \alpha_I, X_\infty \rangle$ switches sign.  Since the basis of positive co-roots determined from $X_\infty$ changes, the definition of the anti-fundamental Weyl chamber changes, and hence the \tHooft charge will no longer be in the anti-fundamental chamber.  We then expect that the moduli space will have a positive dimension.  

To determine the change in the index, we note that the old and new bases of simple co-roots are related by a Weyl reflection about the root $\alpha_I$, and hence the components of the \tHooft charge with respect to the new basis can be obtained by applying the same Weyl transformation to $P^-$.  However, we know that the  $\sum_{\alpha \in \Delta} | \langle \alpha, P \rangle|$ term of \eqref{adindex} is invariant under Weyl reflections, so the change in the index is again given by \eqref{DeltaIndexad}, with $\gm = P^-$.  

We can give an expression for this in terms of the $\mathfrak{u}(N)$ charges, $p^I$, of the \tHooft defect, that will be useful for comparing with the brane result.  We recall from \eqref{gentHooftcharge} that the generic \tHooft charge has the form
\begin{equation}\label{Pofp}
P = \sum_{J=1}^{N-1} (p^J - \bar{p}) H_{J(N-1)}~,
\end{equation}
where $H_{J(N - 1)} = H_J + \cdots + H_{N-1}$ is a co-root and $\bar{p} = \frac{1}{N} \sum_{I=1}^N p^I$.  Saying $P = P^-$ corresponds to an ordering $p_1 \leq p_2 \leq \cdots \leq p_N$.  Then applying \eqref{genrootgenP} in the case of a simple root $\alpha_I = \alpha_{II}$, we find $\langle \alpha_{I}, P^- \rangle = p^I - p^{I+1} = - | p^I - p^{I+1} |$.  Therefore the change in the dimension of the moduli space is
\begin{equation}\label{puretHooftwc}
\Delta_I \dim{\fMM} = 2 \Delta_I \ind{L} = -4 \langle \alpha_I, \gm \rangle = -4 \langle \alpha_I, P^- \rangle = 4 |p^I - p^{I+1}|~.
\end{equation}
This is also the dimension of the moduli space after the wall, since the dimension of the moduli space before the wall is zero.  

\begin{figure}
\begin{center}
\includegraphics{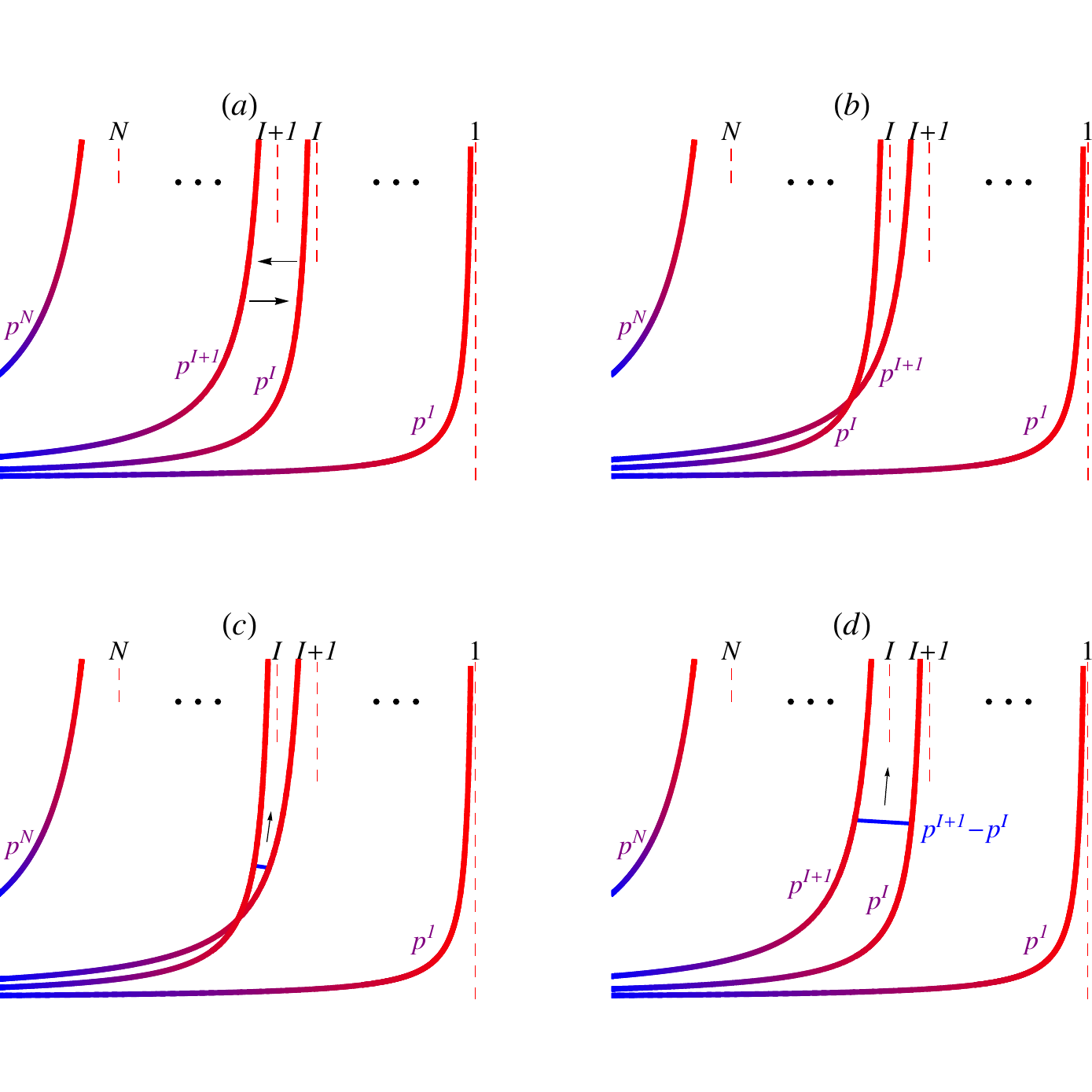}
\caption{Wall-crossing for a Cartan-valued singular monopole in $\mathfrak{g} = \mathfrak{su}(N)$ theory. $(a)$: The initial configuration with \tHooft charge \eqref{Pofp}, with $P = P^-$ such that $p^1 \leq p^2 \leq \cdots \leq p^N$.  We interchange the positions of the $I^{\rm th}$ and $(I+1)^{\rm th}$ brane as indicated by the arrows, resulting in configuration $(b)$, where the branes now intersect.  $(c)$: we can view the intersection locus as the position of some D1-strings of infinitesimal length; (in the figure we displace the D1-strings from the locus slightly for clarity). $(d)$: the infinitesimal D1-strings can be moved out of the throat region as indicated by the arrows, resulting in a defect with the original charge $P^-$, and $p^{I+1} - p^I$ finite length mobile D1-strings.}
\label{fig14}
\end{center}
\end{figure}

The wall-crossing we have just described corresponds to the brane motion depicted in Figure \ref{fig14}.  If we start with a pure \tHooft defect of charge $P = P^-$ in $G = PSU(N)$ gauge theory, this corresponds to the brane configuration of Figure \ref{fig14}-$(a)$, where for all $J = 1,\ldots,N-1$, the throat of the $J^{\rm th}$ D3-brane stays inside and does not cross the throat of the $(J+1)^{\rm th}$ D3-brane.  Sending $\langle \alpha_I, X_\infty \rangle$ through zero corresponds to moving the $I^{\rm th}$ D3-brane to the left and past the $(I+1)^{\rm th}$ brane.  The resulting configuration is depicted in Figure \ref{fig14}-$(b)$.  It still represents a Cartan-valued solution, but with an \tHooft charge that is no longer in the closure of the anti-fundamental Weyl chamber.  We can see this from the fact that the brane worldvolumes now intersect.  As we have argued in the previous section, the brane intersection is a locus from which mobile D1-strings can be extracted, effecting in the process a local Weyl transformation of the \tHooft charge that sends it back to the anti-fundamental Weyl chamber.  This is depicted in Figures \ref{fig14}-$(c),(d)$.  The number of mobile D1-branes that can be extracted is precisely $| p^I - p^{I+1} | = p^{I+1} - p^I$ and in this example they account entirely for the quaternionic dimension of the new moduli space, matching onto \eqref{puretHooftwc} exactly.

With this example and the smooth monopole case of the previous subsection understood, it is easy to describe wall-crossing in the general situation.  Suppose that in addition to the defect with charge $P$, \eqref{Pofp}, we have some numbers $k^I$ of smooth fundamental monopoles of each type $I$.  (This is depicted as a brane diagram with semi-infinite D1-strings for the defect in Figure \ref{fig7}.)  There is no loss of generality in taking the initial defect charge to be in the closure of the anti-fundamental Weyl chamber.  If it is not we make a local Weyl transformation that puts it there, extracting the appropriate numbers of smooth monopoles in the process.  Then we redefine the $k^I$ to include these monopoles.

Now we decrease $\langle \alpha_I, X_\infty \rangle$ and send it through zero, for some simple root $\alpha_I$.  By the same arguments as in the previous example, this has the effect of conjugating the \tHooft charge by a Weyl reflection, and therefore the contribution of the $\sum_{\alpha \in \Delta} |\langle \alpha, P \rangle |$ term to the index is the same before and after the wall.  Therefore the change in the index is still given by \eqref{DeltaIndexad}, and all we need to know is the asymptotic magnetic charge of the initial configuration.  This was written down in in \eqref{gmgenconfig} and is simply the \tHooft charge, \eqref{Pofp}, plus the contribution from the mobile D1-strings, $\sum_I k^I H_I$.  Since the change in the index, and hence dimension, is linear in $\gm$, we get the sum of \eqref{puretHooftwc} and \eqref{smoothDdim} with $m^I \to k^I$ when the new moduli space is non-empty.  Furthermore the criterion for the new moduli space to be non-empty is that the numbers of mobile D1-strings all be non-negative: $k_{\rm new}^J \geq 0, \forall J$.  The only $k^J$ that changes value, and hence could cause a violation of this condition, is $k^I$.  $k_{\rm new}^I$ receives a contribution from both the old $k^J$ and the new mobile D1-strings extracted from the defect.  The former contribution is $k^{I+1} + k^{I-1} - k^I$ as in the smooth case, while the latter contribution is $|p^I - p^{I+1}|$ as in the previous example.  Hence we arrive at
\begin{align}
& \Delta_I \dim{\fMM} = 4(k^{I+1} + k^{I-1} - 2 k^I) + 4 |p^I - p^{I+1}|~, \cr
& \qquad \qquad \qquad \quad ~\, \textrm{if~} k_{\rm new}^I \equiv k^{I+1} + k^{I-1} + |p^I - p^{I+1}| - k^I \geq 0~, \cr
& \fMM^{\rm new} = \varnothing~, \qquad \textrm{if~} k_{\rm new}^I < 0~.
\end{align}

This result matches perfectly with the brane picture, which is a ``superposition'' of Figure \ref{fig13}, with $m^I \to k^I$, and Figure \ref{fig14}.  The mobile D1-strings that are present in the initial configuration behave just like the mobile D1-strings of Figure \ref{fig13}, while the defect behaves as in Figure \ref{fig14}.  The condition for the new moduli space to be non-empty is also easy to understand; it is again the requirement that there be no anti-D1's left over at the end of the process.  $k^I$ anti-D1's are created from the wall-crossing of the initial $k^I$ D1-strings, as depicted in \ref{fig13}.  Now, in addition to the $k^{I+1} + k^{I-1}$ D1-strings, there are also the $|p^I - p^{I+1}|$ D1-strings extracted from the defect that can be used to annihilate the anti D1's.  Thus as long as $k_{\rm new}^I = k^{I+1} + k^{I-1} + |p^I - p^{I+1}| - k^I \geq 0$, there will be no anti-D1's left over.

\subsection{Wall-crossing of the index for matter in the fundamental}

The index formula \eqref{indLrho} and its wall-crossing behavior \eqref{eq:IndexJump} apply for any representation $\rho$ of $G$.  Suppose we couple the Yang--Mills--Higgs system to fermions transforming in representation $\rho$ by adding to the action
\begin{equation}\label{Smat}
S_{\rm mat} = -\frac{1}{g_{0}^2} \int \ed^4 x \bigg\{ \Psi_{D}^T \left[ i \gamma^\mu \otimes (\pd_\mu + \rho(A_\mu) ) + \gamma^5 \otimes (\rho(X) - i m_\rho) \right] \Psi \bigg\}~.
\end{equation}
Here $\Psi(x)$ is an anti-commuting $\mathbb{C}^4 \otimes V_\rho$-tuple, and $\Psi_D$ is the usual Dirac conjugate, such that $\Psi_{D}^{T} = (\psi^\ast)^T \gamma^0$.  The $\gamma^\mu$ satisfy $\{ \gamma^\mu, \gamma^\nu \} = -2\eta^{\mu\nu}$, and $\gamma^5$ is the $\mathfrak{so}(1,3)$ chirality matrix, $\gamma^5 = i \gamma^0 \gamma^1 \gamma^2 \gamma^3$.  The coefficient of the Yukawa coupling to the Higgs field has been chosen so that the Euclidean Dirac operator $i \hat{\slashed{D}}_\rho$, \eqref{hatDirac}, emerges, but it can also be motivated from the point of view of $\NN = 2$ supersymmetry.  The mass parameter $m_\rho$ is real; analogously to the Higgs field, $X$, it can be viewed as a real slice in a space of complexified mass parameters in the $\NN = 2$ theory.  If the representation $\rho$ is reducible, such that $\rho = \rho_1 \oplus \rho_2 \oplus \cdots$ in terms of irreducible representations, then $m_\rho$ can take a different value on each irreducible component: $m_\rho = m_{\rho_1} \mathbbm{1}_{V_{\rho_1}} \oplus m_{\rho_2} \mathbbm{1}_{V_{\rho_2}} \oplus \cdots$.  The chirality matrix appears in the Yukawa coupling because $X$ is a pseudo-scalar as evidenced by the form of the Bogomolny equation, $B_i = D_i X$, for example.  We can choose a basis for the Minkowski gamma matrices\footnote{Take $\gamma^0 = \diag(\mathbbm{1}_2 , -\mathbbm{1}_2) = \sigma^3 \otimes \mathbbm{1}_2$ and $\gamma^i = i\sigma^2 \otimes \sigma^i$.} such that
\begin{equation}
\gamma^0 \gamma^i = \Gamma^i~, \qquad -i \gamma^0 \gamma^5 = \Gamma^4~,
\end{equation}
in which case the equation of motion for $\Psi$ can be written in the form
\begin{equation}\label{spacetimeDirac}
\left( \begin{array}{c c} i(\pd_0 + \rho(A_0))  & L_{\rho}^\dag - i m_\rho \\ L_{\rho} + i m_\rho & i (\pd_0 + \rho(A_0)) \end{array} \right) \Psi = 0~,
\end{equation}
with $L_\rho$ as in \eqref{Lrho}.  Consider \eqref{spacetimeDirac} in the background of a (singular) monopole configuration $(A_0 = 0, A_i,X)$.  Let $\Psi(x) = e^{i \EE t} \Psi(\vec{x})$.  Then an eigenfunction $\Psi(\vec{x})$ of the Dirac operator $(i \hat{\slashed{D}})_\rho$, \eqref{hatDirac}, with eigenvalue $\lambda$, corresponds to a mode with frequency $\EE = |m_\rho| + \lambda$.  

In theories with $\NN = 2$ supersymmetry, the fermions of \eqref{Smat} will be accompanied by scalar degrees of freedom.  Solutions to $(i \hat{\slashed{D}})_{\rho} \Psi = 0$ play a special role, in that they preserve half of the supersymmetry and saturate a generalization of the Yang--Mills--Higgs BPS bound, that includes a contribution from the matter sector.  This is the classical limit of the exact quantum BPS bound obtained by Seiberg and Witten for $\NN = 2$ gauge theories with matter, \cite{Seiberg:1994aj}.

Geometrically, over each point $[\hat{A}] \in \fMM$, the zero-modes of $L_\rho$ form a vector space of dimension $\ind{L_\rho}$.  These patch together to form a vector bundle over $\fMM$ which is in fact the index bundle of the operator $L_\rho$ \cite{Taubes:1984je,Manton:1993aa}.  Quantum states are represented by $\LL^2$ sections of this bundle in the semiclassical quantization of the Yang--Mills--Higgs theory with matter \cite{Manton:1993aa,Sethi:1995zm,Cederwall:1995bc,Gauntlett:1995fu,Henningson:1995hj}.

Our goal here is to understand the inclusion of matter fermions, \eqref{Smat}, their zero-modes, and their wall-crossing behavior using an appropriate modification of the brane systems we have been studying.  We begin with our system of $N$ D3-branes realizing Yang--Mills--Higgs theory with $\mathfrak{g} = \mathfrak{u}(N)$.  We can obtain degrees of freedom in the fundamental representation by considering open (fundamental) strings that stretch from the D3-branes to a new type of D-brane.  The new D-brane must wrap the same $\mathbb{R}^{1,3}$ as the D3-branes in addition to other directions, and the low energy theory describing the interactions of the 3-3 strings and the new strings should have $S_{\rm YMH} + S_{\rm mat}$ as a consistent truncation.  Using the fact that $S_{\rm YMH} + S_{\rm mat}$ can be embedded into an $\NN = 2$ supersymmetric theory, we are led to consider D7-branes that extend in the $x^\mu$ directions as well as $x^{6,7,8,9}$.  They will be placed at the same $x^5$ position as the D3-branes, ($x^5 = 0$, say), and at some fixed values of $x^4$.

Consider a single D7-brane.  Quantization of 3-7 strings leads to scalars and fermions that fill out an $\NN = 2$ hypermultiplet transforming in the fundamental plus anti-fundamental representation $\rho = {\bf N} \oplus \overline{{\bf N}}$, as well as excited string states that are neglected in the low energy limit.  Supersymmetry dictates the form of the low energy interactions between 3-3 and 3-7 strings; the 3-7 fermions will couple to the Yang--Mills--Higgs sector through $S_{\rm mat}$.  We get a direct sum of the fundamental and anti-fundamental representation because the strings carry an orientation and can either begin or end on the D3-brane.  

The mass parameter takes the form $m_\rho = x_{\rm D7}^4 \mathbbm{1}_{{\bf N}} \oplus (-x_{\rm D7}^4) \mathbbm{1}_{\overline{{\bf N}}})$, where $x_{\rm D7}^4$ gives the displacement of the D7-brane from the origin in the $x^4$ direction, (and where we have set the string length to one).  To see this, recall that the origin was defined to be the position of the D3-branes when the $\mathfrak{u}(N)$-valued vev $X_{\infty}^{\mathfrak{u}(N)} = 0$.  Furthermore the mass $|x_{\rm D7}^4|$ of the lightest 3-7 string mode is given by the length of the string connecting the D3-brane to the D7-brane.  Now  suppose we turn on the vev $X_{\infty}^{\mathfrak{u}(N)}$.  Let $\{ {\bf e}_\mu ~|~ \mu \in \Delta_{\bf N} \}$ denote a basis for the fundamental representation, so that $ \{ {\bf e}_{-\mu} \}$ is a basis for the anti-fundamental, and expand $\Psi = \sum_{\mu} \Psi^{(\mu)} {\bf e}_\mu + \Psi^{(-\mu)} {\bf e}_{-\mu}$.  Then we have $\rho(X_{\infty}^{\mathfrak{u}(N)}) {\bf e}_\mu = -i \langle \mu, X_{\infty}^{\mathfrak{u}(N)} \rangle {\bf e}_\mu$, and it follows that the mass of the fermion $\Psi^{( \pm \mu)}$ is $\pm (x_{\rm D7}^4 - \langle \mu, X_{\infty}^{\mathfrak{u}(N)} \rangle)$.  The absolute value of this quantity should be the length of a fundamental 3-7 string.  Thus we wish to identify the $\{ \langle \mu, X_{\infty}^{\mathfrak{u}(N)} \rangle ~|~ \mu \in \Delta_{{\bf N}} \}$ with the displacements of the D3-branes in $x^4$.  This is completely consistent with our initial discussion in section \ref{sec:branereview} since the $\langle \mu,  X_{\infty}^{\mathfrak{u}(N)} \rangle$ are the diagonal components of $i X_{\infty}^{\mathfrak{u}(N)}$ in the fundamental representation.  As a quick check, the highest weight can be taken as the first fundamental weight, $\mu = \lambda^{I=1}$.  We identify $\langle \lambda^1, X_{\infty}^{\mathfrak{u}(N)} \rangle$ with the position of the right-most D3-brane.  Then the remaining weights of the fundamental representation are
\begin{equation}\label{Nweights}
\Delta_{\bf N} = \left\{ \lambda^1~,~ \lambda^1 - \alpha_1~,~ \lambda^1- \alpha_1 - \alpha_2~,~ \ldots~,~ \lambda^1 - \sum_{I=1}^{N-1} \alpha_I \right\}~.
\end{equation}
Recalling that $\langle \alpha_I, X_{\infty}^{\mathfrak{u}(N)} \rangle$ gives the distance between the $I^{\rm th}$ and $(I + 1)^{\rm th}$ D3-brane, $\langle \lambda^1 - \sum_{J=1}^I \alpha_J,  X_{\infty}^{\mathfrak{u}(N)} \rangle = \langle \lambda^1 - \alpha_{1I}, X_{\infty}^{\mathfrak{u}(N)} \rangle$ will be the $x^4$ position of the $(I+1)^{\rm th}$ D3-brane.

In order to get to the $\mathfrak{su}(N)$ theory we must project out the central $\mathfrak{u}(1)$ degree of freedom.  Recall the canonical isomorphism $\mathfrak{u}(N) \to \mathfrak{u}(1) \oplus \mathfrak{su}(N)$, \eqref{Phicon}, with respect to which we can decompose the adjoint-valued fields, $\hat{A}^{\mathfrak{u}(N)} = \hat{A}^{\mathfrak{u}(1)} + \hat{A}$, where $\hat{A}$ is $\mathfrak{su}(N)$-valued.  In particular for the asymptotic vev, the $\mathfrak{u}(1)$ component is literally the center-of-mass position of the D3-branes:
\begin{equation}
i X_{\infty}^{\mathfrak{u}(1)} = \frac{1}{N} \tr_{\bf N} \left( i X_{\infty}^{\mathfrak{u}(N)} \right) \mathbbm{1}_{\bf N} = \frac{1}{N} \sum_{\mu \in \Delta_{\bf N}} \langle \mu, X_{\infty}^{\mathfrak{u}(N)} \rangle \mathbbm{1}_{\bf N}  \equiv x_{\textrm{D3-cm}}^4 \mathbbm{1}_{\bf N}~.
\end{equation}
Thus we have $\langle \mu, X_{\infty}^{\mathfrak{u}(N)} \rangle = x_{\textrm{D3-cm}}^4 + \langle \mu, X_\infty \rangle$, and so $\langle \mu, X_\infty \rangle$ measures the positions of the D3-branes relative to the center of mass.  When we had only the Yang--Mills--Higgs theory we commented that the $\mathfrak{u}(1)$ degrees of freedom decoupled from the rest and could simply be neglected.  Now, however, we see that they do couple to the matter fermions.  In order to reduce to the $\mathfrak{su}(N)$ theory we should take their dynamics to be trivial. 

\begin{figure}
\begin{center}
\includegraphics{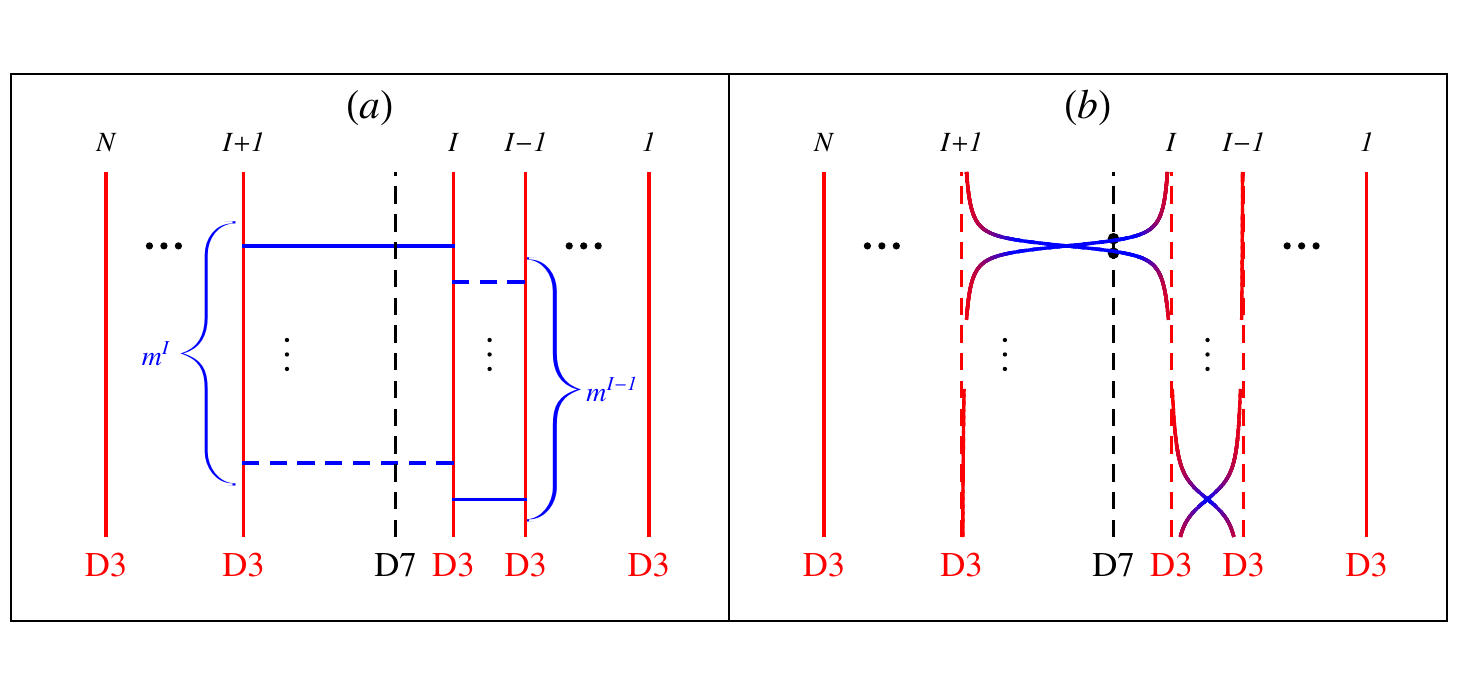}
\caption{$(a)$ Adding a D7-brane to the D1-D3 system representing a generic smooth $\mathfrak{su}(N)$ monopole configuration.  We place the D7-brane at the center of mass position of the D3-branes.  Fundamental strings (not drawn) that are stretched between the D7-brane and the D3-branes give matter in the $\rho = {\bf N} \oplus \overline{{\bf N}}$ representation. $(b)$ We replace the D1-strings with the more precise brane-bending representation of the monopoles.  In order to keep the figure from becoming overly cluttered we represent only the solid D1-strings of $(a)$.  We see that the D7-brane intersects the D3-brane worldvolume.  3-7 strings can have infinitesimal length at these loci and the wavefunctions of the $\LL^2$ modes are localized in their vicinity.  We represent these by black dots.}
\label{fig15}
\end{center}
\end{figure}

Henceforth we focus on the case without defects.  When defects are absent we take $X^{\mathfrak{u}(1)} = X_{\infty}^{\mathfrak{u}(1)}$, $A^{\mathfrak{u}(1)} = 0$.  We choose to place the center of mass position at the location of the D7-brane, $x_{\textrm{D3-cm}}^4 = x_{\rm D7}^4$.  This is a convenience and is done so that the operator, $L_{\rho}^{\mathfrak{u}(N)} -i m_\rho$, that controls the spectrum of the matter fermions in the $\mathfrak{u}(N)$ theory, \eqref{spacetimeDirac}, reduces exactly to the operator we wish to study, $L_{\rho}^{\mathfrak{u}(N)} -i m_\rho = L_{\rho}^{\mathfrak{su}(N)} \equiv L_\rho$.  Thus the brane setup we are considering is Figure \ref{fig15}-$(a)$, and our goal now is to understand the physical mechanism that leads to $\ind{L_\rho}$ zero-modes for the matter fermions represented by 3-7 strings.

Let us recall the form of $\ind{L_\rho}$.  From \eqref{nonnegP1} we see that it can be expressed as
\begin{equation}
\ind{L_\rho} = \sum_{ \mathclap{ \substack{ \mu \in \Delta_\rho \\ \langle \mu, X_\infty \rangle > 0 }} } n_{\rho}(\mu) \langle \mu, \gm \rangle~,
\end{equation}
when \tHooft defects are absent.  With $\rho = {\bf N} \oplus \overline{{\bf N}}$ we simply get double the contribution we would have gotten with $\rho = {\bf N}$.  Furthermore the degeneracies are all one, so
\begin{equation}
\ind{L_{ {\bf N} \oplus \overline{{\bf N}} }} ~=~ 2 \sum_{ \mathclap{ \substack{ \mu \in \Delta_{{\bf N}} \\ \langle \mu, X_\infty \rangle > 0 }} } \langle \mu, \gm \rangle~.
\end{equation}

As is clear from the brane picture, the general case is that the first $I$ weights of \eqref{Nweights} have $\langle \mu, X_\infty \rangle > 0$ while the remaining $N-I$ have $\langle \mu, X_\infty \rangle < 0$, for some $1 \leq I < N$.  Let $\gm = \sum_{K=1}^{N-1} m^K H_K$ and consider $\langle \lambda^1 - \alpha_{1J} , H_K \rangle$.  Using the same techniques as around \eqref{fundtrick} we find 
\begin{equation}
\langle \lambda^1 - \alpha_{1J} , H_K \rangle = \delta_{(J+1)K} - \delta_{JK}~,
\end{equation}
while $\langle \lambda^1, H_K \rangle = \delta_{1K}$.  Thus
\begin{equation}\label{weightscharge}
\langle \lambda^1, \gm \rangle = m^1~, \qquad \langle \lambda^1 - \alpha_{1J}, \gm \rangle = m^{J+1} - m^J ~.
\end{equation}
Summing over $J$ from $1$ to $I-1$, corresponding to the first $I$ weights,
\begin{align}\label{fundindres}
\ind{L_{ {\bf N} \oplus \overline{{\bf N}} }} ~=&~ \langle \lambda^1, \gm \rangle + \sum_{J=1}^{I-1}  \langle \lambda^1 - \alpha_{1J}, \gm \rangle \cr
=&~2 \left( m^1 + (m^2- m^1) + \cdots + (m^I - m^{I-1}) \right) = 2 m^I ~.
\end{align}

Now let us interpret this result in terms of the brane system, Figure \ref{fig15}.  Having the first $I$ weights satisfy $\langle \mu, X_\infty \rangle > 0$, with the remaining weights giving a negative value, corresponds to having the D7-brane between the $I^{\rm th}$ and $(I+1)^{\rm th}$ D3-branes.  The result \eqref{fundindres} suggests that the 3-7 strings corresponding to $\LL^2$-normalizable modes are strings that stretch from the D7-brane to the $I^{\rm th}$ or $(I+1)^{\rm th}$ D3-brane.  The brane picture indeed provides a dynamical mechanism to explain why these strings can have normalizable modes while 3-7 strings ending on other D3-branes do not.  This mechanism is easier to visualize when we represent the monopoles as a localized bending of branes, \ref{fig15}-$(b)$, (and as we discussed around Figure \ref{fig1}).

First note that a 3-7 string ending on a different D3-brane will necessarily pass through one of the two D3-branes adjacent to the D7-brane; thus it can break into a shorter 3-7 string that ends on one of these D3-branes and a 3-3 string.  (The amplitude for this process is captured in the low energy limit by the cubic interaction terms in $S_{\rm mat}$.)  We see why it is dynamically preferable to do this when we consider the bending of the D3-branes.  Notice that the $I^{\rm th}$ or $(I+1)^{\rm th}$ D3-brane will intersect the D7-brane in the vicinity of a fundamental monopole of type $I$.  3-7 strings that stretch between the D7-brane and one of these two D3-branes will be able to shorten their length (and hence reduce their energy) by approaching this intersection locus.  This suggests that the $\LL^2$ modes of 3-7 strings should be thought of as states of the infinitesimal strings connecting the D3-branes and D7-brane in the vicinity of the intersection locus.  

It is harder to argue purely from the brane picture why there are precisely 2 normalizable modes for each monopole of type $I$.  However when the monopoles are well separated we can give a heuristic argument.  In this case the local potential trapping these zero-modes possesses approximate spherical symmetry about the monopole.  One suspects that the wavefunction of the  3-7 fermion zero-mode will be in a singlet state of orbital angular momentum and then the factor of two comes from having a 3-7 string of either orientation.  This suspicion is confirmed by the explicit solution to the Dirac equation in the background of a single $\mathfrak{su}(2)$ monopole \cite{Jackiw:1975fn,Manton:1993aa}.

Having now a clear picture for the brane interpretation of the index formula \eqref{fundindres}, it is easy to understand wall-crossing.  There are now two types of wall-crossing phenomena as we dial $X_\infty$, starting from a generic vector in the fundamental Weyl chamber.  We can encounter a wall of the fundamental Weyl chamber where $\langle \alpha_J, X_\infty \rangle \to 0$ for some simple root.  This wall-crossing corresponds to exchanging the order of the $J^{\rm th}$ and $(J+1)^{\rm th}$ D3-brane and has already been discussed.  However the second type of wall-crossing can occur while keeping $X_\infty$ in the fundamental Weyl chamber and corresponds to $\langle \mu, X_\infty \rangle \to 0$ for some fundamental weight.  In the brane picture this happens when the corresponding D3-brane crosses the D7-brane.  Recall that we have placed the D7-brane at the center of mass position of the D3-branes.  Thus, wall-crossing of this type can only occur for the middle $N-2$ weights (branes).

Suppose that the $I^{\rm th}$ D3-brane crosses the D7-brane so that the D7-brane now lies between the $I^{\rm th}$ and $(I-1)^{\rm th}$ D3-brane.  Denote the $I^{\rm th}$ weight $\mu_I = \lambda^1 - \alpha_{1(I-1)}$.  The change in the index, according to \eqref{eq:IndexJump} will be
\begin{equation}
\Delta_{\mu_I} \ind{L_{ {\bf N} \oplus \overline{{\bf N}} }} ~= - 2 \langle \lambda^1 - \alpha_{1(I-1)} , \gm \rangle = -2 (m^I- m^{I-1} )~,
\end{equation}
since there are no other weights parallel to this weight ,and since $\langle \mu_I, X_\infty \rangle < 0$ after the wall.  In the last step we used \eqref{weightscharge}.  Hence the new index is
\begin{equation}
\ind^{\rm new} L_{ {\bf N} \oplus \overline{{\bf N}} }~= 2 m^I - 2(m^I - m^{I-1}) = 2 m^{I-1}~.
\end{equation}
This result is perfectly consistent with our brane interpretation.  

Indeed, when we take into account the bending of the D3-branes as in Figure \ref{fig15}-$(b)$, we can visualize what happens to the $\LL^2$ modes of the 3-7 strings.  The wavefunctions for these modes should be localized in the vicinity of the D3-D7 intersection.  As the D7-brane position approaches the asymptotic position of the $I^{\rm th}$ D3-brane, these loci move further away from the monopoles until they disappear entirely or move out to infinity, whence the wavefunctions fail to be $\LL^2$.  As the D7-brane position continues past the asymptotic D3-brane position, there is a new set of intersection loci.  The vicinity of these loci is where the new $\LL^2$ wavefunctions are localized.   As the distance between the D7 and D3 continues to increase the loci approach the locations of the $m^{I-1}$ fundamental monopoles of type $I-1$.

This concludes our discussion of wall-crossing for matter in the fundamental representation in the case of smooth monopole configurations.  We expect a similar analysis can be carried out for singular monopole configurations.  However there are a few extra issues to understand concerning the decoupling of the $\mathfrak{u}(1)$ degrees of freedom, and we will not work out the details here.

\section{Conclusions}

In this paper we studied brane realizations of singular monopoles---that is, solutions to the Bogomolny equation with singularities corresponding to the insertion of \tHooft defects---using systems of D3- and D1-branes.  We showed how the brane systems provide physical intuition for the dimension formula, by identifying the motion of finite length D1-string segments with motion on moduli space.  In order to interpret the dimension formula in this way, we found that it is important to represent \tHooft defects in terms of BIon spikes, rather than as semi-infinite D1-strings.  In other words it is important to take into account the effects of brane bending.

We showed how one can make contact with the process of monopole bubbling by considering certain brane motions.  More work needs to be done to gain a fuller understanding of both the global structure and singularity structure of the moduli spaces $\fMM(P;\gm;X_\infty)$.

Also in this paper we considered only the case of $\mathfrak{g} = \mathfrak{su}(N)$.  Smooth monopoles for $\mathfrak{so}$ and $\mathfrak{sp}$ Lie algebras have been considered in the context of branes \cite{Elitzur:1998ju,Ahn:1998ku}, and it would be interesting to generalize those constructions to the case of singular monopoles.

\section*{Acknowledgements}

We thank Sergey Cherkis, Kimyeong Lee, and Edward Witten for helpful discussions.  GM and ABR thank the Aspen Center for Physics  and the NSF grant \#1066293 for hospitality during the completion of this work.  DVdB is partially supported by TUBITAK grant 113F164 and BAP grant 13B037, GM is supported by the U.S. Department of Energy under grant DE-FG02-96ER40959, and ABR is supported by the Mitchell Family Foundation.  GM also thanks the KITP for hospitality during the final phases of the 
preparation of this paper (National Science Foundation under Grant No. NSF PHY11-25915); ABR thanks the NHETC at Rutgers University.

\bigskip


\bibliographystyle{utphys}
\bibliography{MRVdimP2}

\end{document}